\documentclass{jpc} 


\usepackage{natbib}
\setcitestyle{authoryear}
\usepackage[ruled]{algorithm2e}

\usepackage{dirtytalk}
\usepackage{amsfonts,amssymb,mathrsfs,mathdots,mathtools}
\usepackage{dsfont}
\usepackage{subcaption}

\newcommand{\cD}{\ensuremath{\mathcal{D}}}

\newcommand{\cP}{\ensuremath{\mathcal{P}}}

\newcommand{\cU}{\ensuremath{\mathcal{U}}}

\newcommand{\cX}{\ensuremath{\mathcal{X}}}
\newcommand{\cY}{\ensuremath{\mathcal{Y}}}
\newcommand{\cZ}{\ensuremath{\mathcal{Z}}}

\newcommand{\Ren}{R\'enyi }

\DeclarePairedDelimiter\abs{\lvert}{\rvert}%

\DeclareMathOperator*{\argmax}{arg\,max}
\DeclareMathOperator*{\argmin}{arg\,min}
\DeclareMathOperator*{\esssup}{ess\,sup}

\newcommand{\ubar}[1]{\text{\b{$#1$}}}

\allowdisplaybreaks

\begin{document}

\title[Rethinking Disclosure Prevention with Pointwise Maximal Leakage]{Rethinking Disclosure Prevention with Pointwise Maximal Leakage}

\author[S.~Saeidian]{Sara Saeidian}	
\address{KTH Royal Institute of Technology, 100 44 Stockholm, Sweden}	
\email{saeidian@kth.se}  

\author[G.~Cervia]{Giulia Cervia}	
\address{IMT Nord Europe, Centre for Digital Systems, F-59000 Lille, France}	
\email{giulia.cervia@imt-nord-europe.fr}  

\author[TJ.~Oechtering]{Tobias J. Oechtering}	
\address{KTH Royal Institute of Technology, 100 44 Stockholm, Sweden}	
\email{oech@kth.se}

\author[M.~Skoglund]{Mikael Skoglund}	
\address{KTH Royal Institute of Technology, 100 44 Stockholm, Sweden}	
\email{skoglund@kth.se}





\begin{abstract}
\noindent This paper introduces a paradigm shift in the way privacy is defined, driven by a novel interpretation of the fundamental result of Dwork and Naor about the impossibility of absolute disclosure prevention. We propose a general model of utility and privacy in which utility is achieved by disclosing the value of \emph{low-entropy} features of a secret $X$, while privacy is maintained by hiding the value of \emph{high-entropy} features of $X$. Adopting this model, we prove that, contrary to popular opinion, it is possible to provide meaningful \emph{inferential} privacy guarantees. These guarantees are given in terms of an operationally-meaningful information measure called \emph{pointwise maximal leakage} (PML) and prevent privacy breaches against a large class of adversaries regardless of their prior beliefs about $X$. We show that PML-based privacy is compatible with and provides insights into existing notions such as differential privacy. We also argue that our new framework enables highly flexible mechanism designs, where the randomness of a mechanism can be adjusted to the entropy of the data, ultimately, leading to higher utility. 
\end{abstract}

\maketitle

\section{Introduction}
Research in the privacy domain continuously evolves as novel notions of privacy aim to address challenges emerging in applications of data science. Arguably, one of the most successful notions of privacy is \emph{differential privacy}~\citep{dworkCalibratingNoiseSensitivity, dwork2014algorithmic}. Introduced by \cite{dworkCalibratingNoiseSensitivity}, differential privacy guarantees that an individual partaking in a data processing scheme will not face substantially increased risks due to their participation. This guarantee is achieved by ensuring that the outcome of the data processing is not much affected by whether or not each person participates. On the other hand, differential privacy, by design, does not rule out the possibility of privacy violations by \emph{association}. That is, an adversary can still exploit \emph{correlations} among pieces of data to uncover sensitive information about an individual from the outcome of a differentially private mechanism. To account for potential privacy violations by association, \citet{tschantzSoKDifferentialPrivacy2020a} argue that differential privacy should be understood as a \emph{causal} property of an algorithm. That is, differential privacy simply ensures that an algorithm produces similar outputs when supplied with inputs that differ in a single parameter. From the causal standpoint, (an unintended) inference about an individual is considered to be a privacy breach only if it is specifically caused by the inclusion of the individual's information in a dataset~\citep{kifer2022bayesian}.  

On the other hand, the above causal interpretation no longer applies if we adopt a Bayesian perspective and assume that databases are sampled from an underlying probability distribution. In particular, several works argue that from the Bayesian point of view, differential privacy either (implicitly) assumes a product distribution on the database or restricts itself to \emph{informed adversaries}~\citep{kiferNoFreeLunch2011, kifer2012rigorous, he2014blowfish, liu2016dependence, yangBayesianDifferentialPrivacy2015, liMembershipPrivacyUnifying2013, zhu2014correlated}.\footnote{An informed adversary knows all the entries in a database except for one~\citep{dworkCalibratingNoiseSensitivity}.} These works usually provide examples and attack scenarios involving databases containing highly correlated data points and then argue that differential privacy falls short of providing sufficient protection in these cases. For instance, \citet{kiferNoFreeLunch2011} give an example about a medical database in which Bob's data is perfectly correlated with the data of a large number of other patients. Then, they argue that the Laplace mechanism~\citep{dworkCalibratingNoiseSensitivity} does not provide sufficient protection in this case since the effect of Bob's data is amplified by the other data points. These works then often focus on developing tools to guarantee privacy particularly in the case of correlated datasets~\citep{zhu2014correlated, liu2016dependence}. 

A privacy guarantee that can rule out the possibility of privacy breaches due to association must be \emph{inferential} in nature, that is, it must ensure that an adversary's knowledge about the world after interacting with a mechanism does not change much from her prior knowledge.\footnote{We call a privacy notion \emph{inferential} if it is defined by comparing an adversary's posterior knowledge with her prior knowledge. This includes definitions such as maximal leakage~\citep{alvim2014additive, issaOperationalApproachInformation2020}, pointwise maximal leakage~\citep{saeidian2023pointwise_it}, and (local) information privacy~\citep{du2012privacy,jiang2021context} but excludes frameworks that simply assume an underlying distribution on the data, e.g., Pufferfish privacy~\citep{kiferPufferfishFrameworkMathematical2014} or Bayesian differential privacy~\citep{yangBayesianDifferentialPrivacy2015}.} However, inferential guarantees are generally considered to be impossible to achieve by the negative results of \cite{dworkDifficultiesDisclosurePrevention2010a} and \cite{kiferNoFreeLunch2011} (see also \citep[Sec. 7.1]{kifer2022bayesian}). Particularly, \citet{dworkDifficultiesDisclosurePrevention2010a} prove that (under certain assumptions), no mechanism providing \emph{non-trivial utility} can prevent disclosures against adversaries who may possess \emph{auxiliary information} about a secret $X$. This is because an adversary may exploit auxiliary information to disclose more information than what a mechanism intended to release. As an illustrating example, suppose each person's exact height is a secret, and consider a database containing height measurements of people with different nationalities. Assume that the average heights of women of different nationalities are released. Then, an adversary who observes the released values and has the auxiliary information \say{Terry Gross is two inches shorter than the average Lithuanian woman} learns Terry Gross' exact height~\citep{dworkDifficultiesDisclosurePrevention2010a}. Here, if we adopt an inferential view of privacy naively we may conclude that Terry Gross' privacy rights are violated. 

\subsection{Overview and Contributions}
At a high level, \citet{dworkDifficultiesDisclosurePrevention2010a} demonstrate that to provide utility a mechanism necessarily has to disclose some information. To account for this result, differential privacy was designed to distinguish between data that is part of a dataset $X$ and data that is not part of $X$ but correlated with it, where the former is protected but the latter may be disclosed. We call this distinction the \emph{in/out dichotomy}. In this paper, we argue that the in/out dichotomy is not the only way of distinguishing between information that should be protected through privacy guarantees and information that may be disclosed. In particular, we present an alternative distinction termed the \emph{local/global dichotomy}. The concept of the local/global dichotomy yields a fresh perspective on privacy which is compatible with the Bayesian view rather than the causal one required by differential privacy. 

The key to enabling our paper's findings is a fundamental and application-agnostic examination of what constitutes meaningful privacy and what we may consider as utility in privacy-preserving analytics. Roughly speaking, we define privacy as the ability of a mechanism to hide properties that are unique to each realization of the secret $x \in \mathcal X$. We call these properties \emph{local} features of $X$. Conversely, we define utility as the ability of a mechanism to disclose properties of the entire population of $X$, that is, properties that $X$ satisfies with high probability. These properties are called \emph{global} features of $X$. We formally characterize local and global features of $X$ using the concept of \emph{min-entropy} of a probability distribution. Then, we prove that a recently proposed information measure called \emph{pointwise maximal leakage} (PML)~\citep{saeidian2023pointwise_it,saeidian2023pointwise_isit} can be used to protect local features of $X$ but disclose global features of it. That is, we prove that PML achieves privacy according to the local/global dichotomy. Most notably, we establish that to avoid privacy breaches, it is sufficient to make privacy guarantees based solely on assumptions about the true data-generating distribution, and without the need to assess the subjective information leaked to each adversary. It should be emphasized that PML measures information leakage through posterior-to-prior comparisons. Hence, our results indicate that, contrary to popular belief, an inferential perspective on privacy is not all at odds with the results of~\citet{dworkDifficultiesDisclosurePrevention2010a}.

Since PML and the local/global dichotomy introduce a new perspective on privacy, we are motivated to ask if existing \emph{indistinguishability}-based definitions~\citep{dworkCalibratingNoiseSensitivity} can be understood and discussed from our inferential viewpoint. We demonstrate that (pure) differential privacy~\citep{dworkCalibratingNoiseSensitivity} and free-lunch privacy~\citep{kiferNoFreeLunch2011} admit several equivalent formulations in terms of PML. These formulations also show that existing privacy-preserving methods such as the Laplace mechanism~\citep{dworkCalibratingNoiseSensitivity} or randomized response~\citep{warner1965randomized} can be used out of the box to achieve privacy in the sense of PML. On top of that, our Bayesian view has the advantage that the calculated privacy parameter takes the data-generating distribution into account. More precisely, we show that when the data-generating distribution has large entropy, the privacy cost associated with the Laplace mechanism can be significantly lower than the differential privacy parameter. 

Our contributions are briefly summarized as follows: 
\begin{itemize}
    \item We characterize privacy and utility and formally define disclosure in terms of min-entropy (Sections~\ref{ssec:privacy_vs_utility} and~\ref{ssec:entropy_disclosure}). Using these concepts, we argue that the results of \cite{dworkDifficultiesDisclosurePrevention2010a} can be retrieved (Proposition~\ref{prop:impossibility_style}) and re-interpreted in our framework in a way that is consistent with the inferential perspective on privacy. 
    
    \item We show that PML provides privacy guarantees according to the local/global dichotomy by relying solely on assumptions about the underlying distribution of the data (Theorem~\ref{thm:pml_entropy}). We also argue that the privacy parameter of a PML-based guarantee is easily interpretable and admits meaningful upper bounds (Section~\ref{ssec:what_is_eps}). 
    
    \item We show how the inferential view can be used to understand the no-free-lunch theorem of~\cite{kiferNoFreeLunch2011} about the inconsistency of utility with privacy under all possible data-generating distributions (Theorem~\ref{thm:finite_capacity}).
    
    \item We show that pivotal definitions such as pure differential privacy and free-lunch privacy admit several equivalent formulations in terms of PML (Theorems~\ref{thm:dp_pml} and~\ref{thm:free_lunch}); hence, they can be interpreted from the inferential standpoint. 
    
    \item We argue that the inferential perspective offers a significantly flexible design paradigm, where it may even be safe to answer highly general questions about the data deterministically (Example~\ref{ex:deterministic}). We also argue that existing mechanisms can be used more efficiently when privacy is guaranteed in the sense of PML. This is because the privacy cost we pay for answering queries can be adjusted to the entropy of the underlying distribution on the data. We demonstrate this in the case of a counting query answered by the Laplace mechanism (Section~\ref{ssec:laplace}). 
\end{itemize}

\section{Preliminaries}
\subsection{Notation and Terminology}
We use uppercase letters to describe random variables and calligraphic letters to describe sets. Specifically, $X$ denotes some data that contains sensitive information (i.e., the \emph{secret}) and takes values in the finite set $\mathcal X$. We use $P_X$ to represent the (true) probability distribution of $X$, $p_X$ to represent the probability mass function (pmf) of $X$, and $\mathrm{supp}(P_X) \coloneqq \{x \in \mathcal{X}: p_X(x) > 0\}$ to represent the support set of $P_X$. Without loss of generality, we assume that $\mathcal X = \mathrm{supp}(P_X)$. Let $\mathcal P_\mathcal X$ denote the set of all distributions with full support on $\mathcal X$. We use $Q_X \in \mathcal P_\mathcal X$ to represent an adversary's (prior) belief about $X$.\footnote{For convenience, we identify adversaries with their prior beliefs.} Note that $Q_X$ may be different from the true distribution $P_X$ on $X$, but we assume that $Q_X$ and $P_X$ are mutually absolutely continuous. We use $q_X$ to denote the pmf of $Q_X$. 

Let $P_{Y \mid X}$ be a \emph{mechanism}, i.e., a conditional probability kernel, that answers queries about $X$ and let $Y$ represent the public query responses. Suppose $Y$ takes values in the set $\mathcal Y$. The set $\mathcal Y$ may be finite (e.g. if $Y$ is the outcome of the randomized response mechanism~\citep{warner1965randomized}) or infinite (e.g. if $Y$ is the outcome of the Laplace mechanism~\citep{dworkCalibratingNoiseSensitivity}). Given $x \in \mathcal X$, we use $p_{Y \mid X=x}$ to denote the density of $P_{Y\mid X=x}$ with respect to a suitable $\sigma$-finite measure on $\mathcal Y$. For example, when $\mathcal Y$ is a countable set then we use the counting measure and when $\mathcal Y$ is a Euclidean space then we use the Lebesgue measure. Similarly, $P_Y$ denotes the distribution of $Y$ induced by $P_{Y \mid X}$ and $P_X$ and $p_Y$ denotes the density of $P_Y$ with respect to a suitable measure on $\mathcal Y$.  

Let $P_{XY}$ denote the joint distribution of $X$ and $Y$. We write $P_{XY} = P_{Y \mid X} \times P_X$ to imply that $p_{XY}(x,y) = p_{Y \mid X =x}(y) p_X(x)$ for all $(x,y) \in \mathcal X \times \mathcal Y$, where $p_{XY}$ is the density of $P_{XY}$ with respect to a suitable $\sigma$-finite measure on $\mathcal X \times \mathcal Y$. Furthermore, we use $P_{Y} = P_{Y \mid X} \circ P_X$ to denote marginalization over $X$, i.e., to imply that ${p}_{Y}(y) = \sum_{x \in \mathcal X}  p_{Y \mid X=x}(y)  p_X(x)$ for all $y \in \mathcal Y$. These notations can be extended to more than two random variables in a natural way. 

Given random variables $U$, $X$ and $Y$, we say that the Markov chain $U - X - Y$ holds if $U$ and $Y$ are conditionally independent given $X$, that is, if $P_{UY \mid X} = P_{U \mid X} \times P_{Y \mid X}$. This implies that $Y$ depends on $U$ only through $X$ and vice versa. We call a $U$ satisfying the Markov chain $U-X-Y$ an \emph{attribute} or \emph{feature} of $X$, which is induced by the probability kernel $P_{U \mid X}$. Following the terminology of~\citet[p. 96]{dworkDifficultiesDisclosurePrevention2010a}, we call $P_{U \mid X}$ a piece of \emph{auxiliary information} which describes how $U$ depends on the secret $X$. We assume that $U$ takes values in a finite set $\mathcal U$. 

Finally, given a positive integer $n$, $[n] \coloneqq \{1, \ldots, n\}$ describes the set of all positive integers smaller than or equal to $n$, and $\log(\cdot)$ denotes the natural logarithm.

\subsection{Min-entropy and Rényi Divergence of Order Infinity}
We use \emph{min-entropy}, i.e., R\'enyi entropy of order infinity~\citep{renyi1961measures}, as a measure of the uncertainty of a probability distribution.
\begin{defi}[Min-entropy]
Suppose $X$ is a (finite) random variable distributed according to $P_X$. The min-entropy $H_\infty(P_X)$ of $X$ is
\begin{equation*}
    H_\infty(P_X) = - \log \left( \max_{x \in \mathcal X}  p_X(x) \right).
\end{equation*}
\end{defi}
Note that $H_\infty(P_X)$ is maximized when $P_X$ is uniform over $\mathcal X$, and becomes zero when $P_X$ is degenerate, i.e., when $\mathrm{supp}(P_X)$ is a singleton. Henceforth, we use the terms \say{entropy} and \say{min-entropy} interchangeably. 

We now recall the definition of Rényi divergence of order infinity~\citep{renyi1961measures, van2014renyi}, which we then use to define PML.
\begin{defi}[{Rényi divergence of order $\infty$~\cite[Thm. 6]{van2014renyi}}] 
Let $P$ and $Q$ be probability measures on a measurable space. Let $p$ and $q$ denote the densities of $P$ and $Q$ with respect to a dominating $\sigma$-finite measure. The Rényi divergence of order $\infty$ of $P$ from $Q$ is 
\begin{equation*}
    D_\infty(P \Vert Q) = \log \left(\esssup_{P} \frac{p}{q}\right),
\end{equation*}
where $\esssup_{P} f = \sup \{c \in \mathbb R \colon P(f > c) > 0\}$ for all measurable functions $f$. 
\end{defi}

\subsection{Pointwise Maximal Leakage}
Pointwise maximal leakage (PML)~\citep{saeidian2023pointwise_it,saeidian2023pointwise_isit} is an operationally meaningful\footnote{Operationally meaningful information measures are those whose expressions arise naturally while studying specific problems, thus, alleviating the need for the ex-post justification of these quantities. An example of an operationally meaningful measure is mutual information, which determines the maximum rate of reliable communication over a memoryless channel.} privacy measure that quantifies the amount of information leaking about a secret random variable $X$ to a single outcome of a mechanism $P_{Y \mid X}$. \citet{saeidian2023pointwise_it} defined PML by considering two different threat models: the \emph{randomized function view} and the \emph{gain function view}. According to the randomized function view (first introduced by~\cite{issaOperationalApproachInformation2020}), PML is defined as the largest increase in the posterior probability of correctly guessing the value of an arbitrary attribute of $X$ compared to the prior probably of correctly guessing the value of that attribute. That is, the adversary of this model is assumed to possess all possible auxiliary information about $X$, making PML a particularly suitable privacy notion for discussing the results of~\cite{dworkDifficultiesDisclosurePrevention2010a}. Moreover, according to the gain function view (first introduced by~\cite{alvim2012measuring}), PML is defined as the largest increase in the posterior expected gain of an adversary compared to her prior gain. \citet[Thm. 2]{saeidian2023pointwise_it} then prove that these two definitions are equivalent and both yield a simple expression in terms of the \Ren divergence of order $\infty$. It is worth emphasizing that these threat models make explicit the type of privacy captured by PML; thus, unlike most other definitions, PML need not rely on natural-language descriptions of a mathematical quantity to be interpreted. Furthermore, PML satisfies a post-processing inequality and increases (at most) linearly under composition~\cite[Lemma 1]{saeidian2023pointwise_it}. 

Below, we define PML and conditional PML which are used extensively in the paper.

\begin{defi}[Pointwise maximal leakage~{\cite[Def. 1]{saeidian2023pointwise_it}}]
\label{def:randomized_function_view}
Suppose $X$ is a random variable on the finite set $\cX$, and $Y$ is a random variable on the set $\cY$ induced by a mechanism $P_{Y \mid X}$. Let $P_X$ be a distribution on $\cX$, and let $P_{XY} = P_{Y \mid X} \times P_X$ denote the joint distribution of $X$ and $Y$. The pointwise maximal leakage from $X$ to $y \in \cY$ is defined as
\begin{equation}
\label{eq:pml_u_sup}
    \ell_{P_{XY}}(X\to y) \coloneqq \log \sup_{P_{U \mid X}} \frac{\sup_{P_{\hat U \mid Y}} \mathbb P \left(U=\hat U \mid Y=y \right)}{\max_{u \in \cU} p_U(u)},
\end{equation}
where $U$ and $\hat U$ are random variables on a finite set $\cU$, $P_{U} = P_{U \mid X} \circ P_X$, and the Markov chain $U-X-Y-\hat U$ holds.
\end{defi}

Definition~\ref{def:randomized_function_view} can be explained as follows. Let $U$ be an arbitrary feature of $X$ induced by a kernel $P_{U \mid X}$. Essentially, $P_{U \mid X}$ is a piece of auxiliary information that describes how $U$ depends on $X$. To measure the information leakage associated with the released outcome $y$, \eqref{eq:pml_u_sup} compares the probability of correctly guessing $U$ having access to $y$ with the probability of correctly guessing $U$ without access. The probability of correctly guessing $U$ with access to $y$ is calculated using the optimal estimator $\hat U$, obtained by taking the supremum over $P_{\hat U \mid Y}$. Conversely, the probability of correctly guessing $U$ without access is simply $\max_{u \in \mathcal{U}} p_U(u)$.

Note that by Definition~\ref{def:randomized_function_view}, PML is robust against arbitrary auxiliary information. This is because PML is defined by measuring the largest posterior-to-prior change across all $U$'s correlated with $X$ represented by the supremum over $P_{U \mid X}$. 

Recognizing the complexity of~\eqref{eq:pml_u_sup}, \citet{saeidian2023pointwise_it} established a simpler expression for PML. 
\begin{thm}[{\cite[Thm. 1]{saeidian2023pointwise_it}}]
\label{thm:pml}
Let $P_{XY}$ be a distribution on the set $\mathcal X \times \mathcal Y$ with the marginal distribution $P_X$ on $\mathcal X$. The pointwise maximal leakage from $X$ to $y \in \mathcal Y$ is\footnote{To be able to define PML for all $y \in \mathcal Y$, we use the convention that $P_{X \mid Y=y} = P_X$ if $p_Y(y) =0$. That is, conditioning on outcomes with density zero equals no conditioning.}
\begin{equation*}
    \ell_{P_{XY}}(X \to y) = D_\infty(P_{X \mid Y=y} \Vert P_X),
\end{equation*}
where $P_{X \mid Y=y}$ denotes the posterior distribution of $X$ given $y \in \mathcal Y$. 
\end{thm}

When the joint distribution used to measure the information leakage is clear from context, we do not specify it as a subscript and write $\ell(X \to y)$. Note that PML is non-negative and bounded above by $- \log \left( \min_{x \in \mathcal X} P_X(x) \right)$. It also satisfies a pre-processing inequality indicating that a mechanism leaks less information about attributes of $X$ compared to $X$ itself~\cite[Lemma 1]{saeidian2023pointwise_it}. Formally, if the Markov chain $Z - X - Y$ holds, then $\ell(Z \to y) \leq \ell(X \to y)$ for all $y \in \mathcal{Y}$.

To quantify the information leakage in the presence of side information \citet{saeidian2023pointwise_it} also define a conditional form of PML. Side information can be modeled as the outcome of a random variable $Z$ correlated with $X$ and $Y$. For instance, when $X$ is a database, side information can represent a subset of the database known to an adversary already before interacting with the mechanism. 

\begin{defi}[Conditional PML~{\cite[Def. 3]{saeidian2023pointwise_it}}]
\label{def:cond_pml}
Let $P_{XY \mid Z}$ denote the conditional distribution of $X$ and $Y$ given $Z$, where $Z$ is a random variable on the set $\cZ$. The conditional pointwise maximal leakage from $X$ to $y \in \mathcal Y$ given $z \in \cZ$ is 
\begin{equation*}
    \ell_{P_{XY \mid Z}}(X \to y \mid z) =  D_\infty(P_{X \mid Y=y, Z=z} \Vert P_{X \mid Z=z}). 
\end{equation*}
\end{defi}

\citet{saeidian2023pointwise_it} define several privacy guarantees by restricting PML in various ways. The simplest definition, called $\epsilon$-PML, bounds the information leaking through the mechanism by $\epsilon \geq 0$ with probability one. Here, we extend the definition of $\epsilon$-PML to encompass scenarios where $P_X$ is not precisely known, but is assumed to belong to a subset of $\mathcal P_\mathcal X$.

\begin{defi}[$(\epsilon, \mathcal P)$-PML]
Suppose $X$ is distributed according to $P_X \in \mathcal P \subseteq \mathcal P_\mathcal X$. Given $\epsilon \geq 0$, we say that the mechanism $P_{Y \mid X}$ satisfies $(\epsilon, \mathcal P)$-PML if 
\begin{equation*}
    P_Y\left(\left\{y \in \mathcal Y : \ell_{P_{Y \mid X} \times P_X}(X \to y) \leq \epsilon \right\} \right) = 1,
\end{equation*}
for all $P_X \in \mathcal P$, or equivalently, if 
\begin{equation*}
    \sup_{P_X \in \mathcal P} \, D_\infty(P_{Y \mid X} \times P_X \Vert P_Y \times P_X) \leq \epsilon.
\end{equation*}
\end{defi}
For simplicity, we assume that the density $p_{Y \mid X=x}(y)$ is continuous on $\mathcal Y$ for all $x \in \mathcal X$.\footnote{See~\cite[Remark 3.15]{rudin86real} for a discussion on replacing the essential supremum by the actual supremum of a function.} In this case, $P_{Y \mid X}$ satisfies $(\epsilon, \mathcal P)$-PML if
\begin{equation*}
    \sup_{P_X \in \mathcal P} \; \sup_{y \in \mathcal Y} \; \ell_{P_{Y \mid X} \times P_X}(X \to y) \leq \epsilon.
\end{equation*}

The choice of an appropriate set of priors $\mathcal{P}$ depends on the specific information available in each application and the corresponding privacy requirements. For instance, as discussed in Section~\ref{ssec:laplace}, while analyzing the counting query, the prior can encompass bounds on the probability that a predicate is satisfied. More generally, when the data has a bounded domain, the prior could incorporate upper and lower bounds on the range of the data. Extending further, one may consider families of distributions, such as sub-Gaussian distributions, distributions that have a well-defined moment-generating function, or distributions with finite $k$-th order moments. 

\subsection{Leakage Capacity}
Now, we define the notion of the \emph{leakage capacity} of a mechanism which, according to \cite[Thm. 14]{issaOperationalApproachInformation2020}, describes the largest amount of information that can leak through a mechanism $P_{Y\mid X}$. 
\begin{defi}[Leakage Capacity]
\label{def:leakage_capacity}
The leakage capacity of a mechanism $P_{Y \mid X}$ is
\begin{equation*}
    C(P_{Y \mid X}) \coloneqq \log \; \sup_{y \in \mathcal Y}\; \max_{x,x' \in \mathcal X} \frac{p_{Y \mid X=x}(y)}{p_{Y \mid X=x'}(y)}. 
\end{equation*}
\end{defi}
Note that $C(P_{Y \mid X})$ is infinite if there exists $(x,y) \in \mathcal X \times \mathcal Y$  such that $p_{Y \mid X=x}(y) = 0$ but $p_Y(y) >0$. In~\citep{fernandes2022explaining}, the quantity $\exp \left(C(P_{Y \mid X})\right)$ is called \emph{lift capacity} and is used to establish a connection between a privacy measure called \emph{max-case $g$-leakage} and local differential privacy~\citep{kasiviswanathan2011can,duchi2013local}. Definition~\ref{def:leakage_capacity} is also related to the notion of \emph{indistinguishability}~\citep{dworkCalibratingNoiseSensitivity} as well as differential privacy.  

\begin{thm}[{\cite[Thm. 14]{issaOperationalApproachInformation2020}}]
\label{thm:pml_ldp}
Given a mechanism $P_{Y \mid X}$ it holds that 
\begin{equation*}
    C(P_{Y \mid X}) = \sup_{P_X \in \mathcal P_\mathcal X} \; \sup_{y \in \mathcal Y} \; \ell_{P_{Y \mid X} \times P_X}(X \to y) = \sup_{P_X \in \mathcal P_\mathcal X} \; D_\infty(P_{Y \mid X} \times P_X \Vert P_Y \times P_X).
\end{equation*}
\end{thm}
In Section~\ref{sec:database_guarantees}, we characterize (pure) differential privacy and free-lunch privacy~\citep{kiferNoFreeLunch2011} in terms of PML. These results can be considered as consequences of Theorem~\ref{thm:pml_ldp}. 

\section{Impossibility of Absolute Disclosure Prevention}
\label{sec:dalenius}

Any aspiring inferential privacy framework should first be reconciled with the results of~\cite{dworkDifficultiesDisclosurePrevention2010a} and~\cite{kiferNoFreeLunch2011}, and this is the subject we take up in this section. We mainly discuss the results of \cite{dworkDifficultiesDisclosurePrevention2010a} as they 
correspond more directly to the ideas laid out in this paper, but we also draw connections to \citep{kiferNoFreeLunch2011}.\footnote{We emphasize that \cite{dworkDifficultiesDisclosurePrevention2010a} and \cite{kiferNoFreeLunch2011} prove conceptually different results. Specifically, \cite{dworkDifficultiesDisclosurePrevention2010a} prove that absolute disclosure prevention is impossible due to the auxiliary information that may be available to an adversary, even if we assume a single fixed and publicly known prior distribution. On the other hand, \cite{kiferNoFreeLunch2011} prove that guaranteeing privacy under all possible prior distributions severely restricts utility.}  

\cite{dworkDifferentialPrivacy} proves a fundamental result that marks the beginning of the developments in the area at of differential privacy. This result, dubbed the \emph{impossibility result}, proves that no mechanism providing \say{non-trivial utility} can prevent disclosures against adversaries who may possess arbitrary \emph{auxiliary information} about a secret $X$.\footnote{The impossibility result is somewhat extended in \citep{dworkDifficultiesDisclosurePrevention2010a} and we mostly refer to ideas from this later version.} Roughly speaking, \cite{dworkDifficultiesDisclosurePrevention2010a} demonstrate that an adversary can exploit auxiliary information to make unintended inferences about quantities correlated with $X$, as illustrated by the example about Terry Gross' height. Thus, privacy guarantees that ensure neither $X$ nor any quantity correlated with $X$ is disclosed can be achieved only at the cost of destroying all utility, because it is not feasible to control for the adversary's auxiliary information. 

The impossibility result states that to provide utility, one necessarily has to disclose some information. This raises the question: What information can we (and should we) protect through privacy guarantees, and what information will we inevitably disclose? The answer differential privacy gives to this question is that privacy guarantees should be limited to information that is directly included in $X$. So, when $X$ is a database, the individuals who have contributed their data to the database should be protected, but no such guarantee is provided to individuals whose data may be correlated with $X$ in other ways. That is, a distinction is made between information that is directly included in $X$ and information that is not part of $X$ but may be correlated with it. We call this distinction the \emph{in/out dichotomy}. As the basis for differential privacy, the in/out dichotomy has proved to be a very useful idea for addressing the impossibility result. 

Nevertheless, the in/out dichotomy is not the only way we can distinguish between the information that we protect and the information that we allow to be disclosed. Below, we present an alternative distinction which we call the \emph{local/global dichotomy}. The idea behind the local/global dichotomy is that we protect features of $X$ that have large entropy (i.e., local features) while we allow disclosing features of $X$ with small entropy (i.e., global features).\footnote{These entropies are calculated using the true prior distribution $P_X$ on the data.} This view is motivated by how we define utility and what we consider to be a privacy breach. In particular, we argue that features of the data that capture properties of the population as a whole have small entropy and may be disclosed for the sake of utility, whereas instance-dependent features of the data have large entropy and should remain secret. Then, similar to how differential privacy provides guarantees according to the in/out dichotomy, we show that PML's guarantees are based on the local/global dichotomy. In short, the local/global dichotomy allows us to reconcile the results of ~\cite{dworkDifficultiesDisclosurePrevention2010a} with the guarantees of an effective inferential privacy framework. The main advantage of this view is that \emph{those features of $X$ that may be revealed by the mechanism are exactly those population-level features of the data that we would anyway want to be able to disclose to provide utility.} On top of that, this view is directly applicable to many types of secrets and not just private databases. 

In what follows, we assume that $X$ is any type of data containing sensitive information, for example, a database or a piece of information belonging to a single individual. All the results presented in this section are proved in Appendix~\ref{sec:proofs_impossibility}. 

\subsection{What Is Privacy and What Is Utility?}
\label{ssec:privacy_vs_utility}
Our results and discussions throughout the paper depend crucially on definitions of utility and privacy formulated in terms of entropy. As such, in this subsection, we recall the definitions and assumptions of~\cite{dworkDifficultiesDisclosurePrevention2010a}, in particular, the notions of utility and privacy posited there. We then present our own definitions and assumptions and discuss how they differ from that of~\cite{dworkDifficultiesDisclosurePrevention2010a}. 

Suppose $X$ is distributed according to $P_X\in \mathcal P_\mathcal X$. \citet{dworkDifficultiesDisclosurePrevention2010a} assume that $P_X$ is publicly known, that is, $P_X$ also represents the prior belief of an adversary who interacts with a mechanism $P_{Y \mid X}$. To define utility, \citet{dworkDifficultiesDisclosurePrevention2010a} posit a random variable $U$ satisfying the Markov chain $U-X-Y$ whose value represents the answer to a question posed about $X$. It is assumed that the value of $U$ cannot be \emph{a priori} predicted from its distribution $P_U = P_{U \mid X} \circ P_X$, that is, the entropy $H_\infty (P_U)$ is large. However, to provide utility, the mechanism must either disclose the value of $U$ exactly or allow estimating $U$ with high accuracy, i.e., it is assumed that there exists $y \in \mathcal Y$ such that the entropy $H_\infty(P_{U \mid Y=y})$ is either very small or zero. Furthermore, to define privacy, \citet{dworkDifficultiesDisclosurePrevention2010a} suppose the existence of a random variable $W$ satisfying the Markov chain $W-X-Y$ whose value must remain secret. That is, the value of $W$ must be difficult to guess with or without access to the mechanism, but it is assumed that $W$ has smaller entropy compared to $U$. Formally speaking, $H_\infty(P_W)$ and $H_\infty(P_{W \mid Y=y})$ are both large for all $y \in \mathcal Y$, but $H_\infty(P_W) < H_\infty(P_U)$. It is important to note that the condition $H_\infty(P_W) < H_\infty(P_U)$\footnote{This condition is implied by the lower bound on the entropy of the utility vector in terms of the length of the privacy breach in~\cite[Assumption 1]{dworkDifficultiesDisclosurePrevention2010a}.} is indispensable in the proof of the impossibility result because~\cite{dworkDifficultiesDisclosurePrevention2010a} assume that it is possible to extract enough randomness from $U$ to mask the value of $W$. 

Our setup differs from \citep{dworkDifficultiesDisclosurePrevention2010a} in several key aspects. We let distribution $Q_X \in \mathcal P_\mathcal X$ represent the prior belief of an adversary who observes the outcome of the mechanism. This distribution may or may not be equal to $P_X$, but $P_X$ and $Q_X$ are mutually absolutely continuous. To provide utility, the mechanism $P_{Y \mid X}$ releases some \emph{global} information about the secret $X$, and releasing this information is \emph{not} considered to be a privacy breach. We define global information as the value of any attribute of $X$ that can be accurately predicted by an analyst who knows the true distribution $P_X$ and possibly some auxiliary information but without access to the mechanism $P_{Y \mid X}$. Formally, we posit a Markov chain $U-X-Y$, where $U$ is an attribute of $X$ and the kernel $P_{U \mid X}$ is the analyst's auxiliary information. If $U$ contains global information about $X$, then the entropy $H_\infty(P_U)$ must be \emph{small} since the value of $U$ should be predictable using the distribution $P_U$ alone (where $P_U = P_{U \mid X} \circ P_X$) and without access to $P_{Y \mid X}$. Heuristically, such attributes describe properties of the population of $X$ and are largely instance-independent. Hence, they may be disclosed to provide utility. By contrast, to maintain privacy, we wish to protect instance-dependent and \emph{local} properties of $X$, which are represented by those attributes of $X$ that have \emph{large} entropy. Consider an attribute $W$ of $X$ satisfying the Markov chain $W-X-Y$. If $H_\infty(P_W)$ is large (where $P_W = P_{W \mid X} \circ P_X$), then even an analyst who knows the true underlying distribution $P_X$ and the auxiliary information $P_{W \mid X}$ cannot reliably estimate $W$; hence, it is only through the mechanism $P_{Y \mid X}$ that the value of $W$ can be disclosed. Accordingly, we consider it to be a privacy breach if the value of any high-entropy attribute of $X$ is disclosed. 

The above distinction between high-entropy local features of $X$ and low-entropy global features of it is what was earlier called the local/global dichotomy. This is further illustrated by the examples below, where the second example is inspired by~\citep{kasiviswanathanSemanticsDifferentialPrivacy2015}.
\begin{exa}
\label{ex:good_bad_mean_estimate}
Suppose the database $X= (D_1, \ldots, D_n)$ is i.i.d, where each entry $D_i$ is drawn according to a distribution $P_D$ defined over a finite set of real numbers in the interval $[a,b)$. Our goal is to estimate the expectation $\mu = \mathbb E_{P_D} [D_i]$. We may aim to disclose one of the following two estimates: the quantized sample mean $\hat \mu_1 = q_m \left(\frac{\sum_{i=1}^n D_i}{n} \right)$, or the first row of the database $\hat \mu_2 = D_1$. The quantization $q_m(.)$ can be described as follows: Fix a large integer $m$, and values $c_1, \ldots, c_{m-1}$ satisfying $a = c_0 < c_1 < \ldots < c_m = b$. Let $\mathcal C = \{\frac{c_0 + c_1}{2}, \ldots, \frac{c_{m-1} + c_{m}}{2}\}$. Then, $q_m : [a, b) \to \mathcal C$ denotes a quantizer that maps real numbers in the interval $[c_j, c_{j+1})$ to $\frac{c_j + c_{j+1}}{2}$.\footnote{By the central limit theorem, the sample mean converges in distribution to a Gaussian random variable as $n \to \infty$. Thus, we use the quantization to ensure that the entropy of our estimator remains well-defined as $n \to \infty$. The quantization introduces some bias, which can be made arbitrarily small by taking $m$ sufficiently large.} 

By the law of large numbers, as $n \to \infty$ the sample mean converges in probability to $\mu$; thus, $H_\infty(\hat \mu_1) \to 0$. In contrast, the distribution of $\hat \mu_2$ does not depend on $n$; hence, $\hat \mu_2$ has larger entropy compared to $\hat \mu_1$. Therefore, a mechanism is allowed to disclose the value of $\hat \mu_1$ for the sake of utility but $\hat \mu_2$ must be kept secret for the sake of privacy. 
\end{exa}

The above example also sheds light on Terry Gross' case: If the average height of Lithuanian women is released using a low-entropy accurate estimator with suitable convergence properties (e.g. $\hat \mu_1$), then we do not consider the disclosure of her height as a privacy breach. This is because an adversary who knows the distribution of women's height can predict her height even without access to the mechanism. 

\begin{exa}
An insurance company has access to an i.i.d medical database $X$ of size $n$ and queries it through $P_{Y \mid X}$ to obtain (quantized) relative frequencies $\hat p_s$ and $\hat p_{ns}$ describing the empirical probabilities of developing lung disease for smokers and non-smokers, respectively. Let $p_s$ and $p_{ns}$ denote the true probabilities of developing lung disease for smokers and non-smokers, which can be calculated from the prior distribution $P_X$. If $n$ is large, then the estimates $\hat p_s$ and $\hat p_{ns}$ have small entropies and well-approximate the true probabilities.  

Now, suppose based on $\hat p_s$ and $\hat p_{ns}$ the company draws some conclusions about Bob's probability of developing lung disease, and adjusts his insurance premium accordingly. Assuming that $\hat p_s$ and $\hat p_{ns}$ well-approximate the true probabilities, we do not consider this to be in violation of Bob's privacy (regardless of his participation in the database). This is because the insurance company could have drawn the same conclusions about Bob from the prior $P_X$ even without access to the mechanism.
\end{exa}

In essence, the differences between our setup and~\citep{dworkDifficultiesDisclosurePrevention2010a} stem from the fundamental principle that if an analyst knows the true distribution $P_X$ on the data, then they should be granted no further utility. Interestingly, the local/global dichotomy also allows us to distinguish between \emph{adversarial} and \emph{non-adversarial} analysts. The non-adversarial analyst Alice is only interested in the value of low-entropy attributes of $X$, which reflect properties of the population as a whole. If Alice knows $P_X$, then she gains no further value from interacting with the mechanism $P_{Y \mid X}$. On the other hand, the adversarial analyst Eve even equipped with $P_X$ is motivated to query $X$ through $P_{Y \mid X}$ to uncover the value of high-entropy, instance-dependent, and local features of $X$ which she cannot \emph{a priori} predict, even if she possesses arbitrary auxiliary information. 

\subsection{Entropy-based Disclosure Prevention}
\label{ssec:entropy_disclosure}
Equipped with our definitions of privacy and utility, in this subsection, we state the main results of the paper: that (a) disclosing a piece of information (in the sense of Definition~\ref{def:disclosure}) to one adversary in $\mathcal P_\mathcal X$ is tantamount to disclosing that information to all adversaries in $\mathcal P_\mathcal X$ (Theorem~\ref{thm:ubiq_disclos}), and (b) PML provides privacy guarantees according to the local/global dichotomy (Theorem~\ref{thm:pml_entropy}). In particular, we show that if a mechanism $P_{Y \mid X}$ satisfies $(\epsilon, P_X)$-PML, then it will not disclose the value of any attribute of $X$ with entropy greater than $\epsilon$ to any adversary with prior belief in the set $P_\mathcal X$. Afterward, in the spirit of the impossibility result, we prove that when a mechanism discloses the value of an attribute $U$ of $X$, then it also discloses another attribute of $X$ with smaller prior entropy compared to $U$. Finally, toward the end of this subsection, we discuss \emph{absolute disclosure prevention}, i.e., we examine the condition ensuring that no attribute of $X$ is disclosed by a mechanism.

We begin by formally defining a notion of \emph{disclosure}. Consider an adversary with prior belief $Q_X \in \mathcal P_\mathcal X$, and let $U$ be an attribute of $X$. Then, the adversary's prior belief about $U$ is $Q_U = P_{U \mid X} \circ Q_X$. We may define disclosure as the event that the adversary's belief about $U$ changes after observing an outcome of the mechanism.\footnote {This is often called Dalenius' desideratum in the literature.} That is, disclosure is the event that $Q_{U} \neq Q_{U \mid Y=y}$ for some $y \in \mathcal Y$, where $Q_{U \mid Y=y} = P_{U \mid X} \circ Q_{X \mid Y=y}$ denotes the adversary's posterior belief about $U$ after observing $y$. Thus, disclosure prevention requires that $Y$ and $U$ be independent. Clearly, this is a very stringent requirement and may necessitate the independence of $X$ and $Y$,\footnote{\cite{rassouli2021perfect} show that under certain conditions it is possible to design $P_{Y \mid X}$ such that $Y$ is independent of $U$ but correlated with $X$.} e.g, if $U=X$. Hence, we instead postulate the following weaker but more intuitive definition that also matches the notions of disclosure considered in~\citep{dworkDifficultiesDisclosurePrevention2010a} and \citep{kiferNoFreeLunch2011}.

\begin{defi}[Disclosure]
\label{def:disclosure}
Let $U$ be an attribute of $X$. We say that the mechanism $P_{Y \mid X}$ discloses the value of $U$ to adversary $Q_X \in \mathcal P_\mathcal X$ if $\; \inf_{y \in \mathcal Y} \; H_\infty(Q_{U \mid Y=y}) = 0$.
\end{defi}

Henceforth, we use the terms \say{disclosure} and \say{disclose} in the sense of Definition~\ref{def:disclosure}. The following theorem asserts that, in fact, we do not need to specify to which adversary a piece of information has been disclosed. This is because disclosures are ubiquitous across $\mathcal P_\mathcal X$. 

\begin{thm}[Ubiquity of Disclosures]
\label{thm:ubiq_disclos}
Let $U$ be an attribute of $X$. If the mechanism $P_{Y \mid X}$ discloses the value of $U$ to an adversary $Q_X \in \mathcal P_\mathcal X$, then it also discloses the value of $U$ to all other adversaries in $\mathcal P_\mathcal X$. 
\end{thm}

We now exploit Theorem~\ref{thm:ubiq_disclos} to prove that PML-based privacy guarantees prevent disclosing high-entropy attributes of $X$ to all adversaries in $\mathcal P_\mathcal X$. 

\begin{thm}[Disclosure prevention via PML]
\label{thm:pml_entropy}
Suppose $X$ is distributed according to $P_X$, and let $U$ be an attribute of $X$ with entropy $H_\infty(P_U) > \epsilon$, where $\epsilon \geq 0$ and $P_U = P_{U \mid X} \circ P_X$. If the mechanism $P_{Y \mid X}$ satisfies $(\epsilon, P_X)$-PML, then $P_{Y \mid X}$ will not disclose the value of $U$ to any adversary $Q_X \in \mathcal P_\mathcal X$. 
\end{thm}

The above theorem contains a powerful idea: It states that if we protect the data under its true distribution, then we are simultaneously preventing privacy breaches against all adversaries in $\mathcal P_\mathcal X$. Furthermore, Theorem~\ref{thm:pml_entropy} demonstrates that the two goals of privacy and utility are not inherently at odds with each other. This is because while PML imposes lower bounds on the remaining uncertainty in the value of high-entropy local attributes of $X$, it does not directly restrict the remaining uncertainty in the value of low-entropy global attributes of $X$. Indeed, when the answer to a query describes a feature of $X$ that has very small entropy, it may even be safe to answer it precisely and without any randomness. We give an example of a query answered deterministically in Section~\ref{ssec:what_is_eps}. 

We emphasize two further points. First, Theorem~\ref{thm:pml_entropy} does \emph{not} mean that mechanism $P_{Y \mid X}$ leaks the same amount of information to all adversaries. For example, an attribute $U$ of $X$ that has small entropy (below $\epsilon$) under the true distribution $P_X$ may have very large entropy according to the belief of adversary $Q_X$. In this case, a mechanism that discloses the value of $U$ leaks a large amount of information (possibly, larger than $\epsilon$) to adversary $Q_X$, and this leakage is captured by $\ell_{Q_{XY}}(X \to y)$, where $Q_{XY} = P_{Y \mid X} \times Q_X$. Nevertheless, Theorem~\ref{thm:pml_entropy} asserts that we need not be alarmed by the large value of $\ell_{Q_{XY}}(X \to y)$ because despite this large leakage, adversary $Q_X$ will not be able to infer the value of any local features of $X$. Put differently, while we may use PML \emph{subjectively} to calculate the amount of information leaked to each adversary, the parameter $\epsilon$ of the privacy guarantee should be determined and interpreted \emph{objectively} according to our assumptions about the true underlying distribution on the data. 

Second, in practice, the true data distribution $P_X$ is often unknown to the system designer. To address this uncertainty, the mechanism $P_{Y \mid X}$ should be designed to satisfy $(\epsilon, \mathcal P)$-PML for some carefully chosen set $\cP \subseteq \cP_\cX$ that includes $P_X$. This ensures that Theorem~\ref{thm:pml_entropy} still applies even when the exact distribution is not known. The choice of $\cP$ depends on the specific application and the assumptions made about the data. For instance, if $X$ is a database with independent entries, $\cP$ could represent the set of all product distributions on $X$. This particular scenario relates closely to differential privacy, which we will explore further in Section~\ref{sec:database_guarantees}. Generally, the more assumptions we can make about the data, the smaller and more precise the set $\cP$ becomes, allowing us to design mechanisms with higher utility.

As a converse to Theorem~\ref{thm:pml_entropy}, we now show that when the mechanism $P_{Y \mid X}$ discloses the value of an attribute $U$ of $X$, then we can no longer guarantee privacy for attributes of $X$ with entropies smaller than $H_\infty(P_U)$. In fact, disclosing $U$ inevitably leads to disclosing another attribute of $X$ with a smaller entropy compared to $U$. 

\begin{prop}
\label{prop:impossibility_style}
Suppose $X$ is distributed according to $P_X$. Assume that the mechanism $P_{Y \mid X}$ discloses the value of an attribute of $X$, denoted by $U$. Then, there exists an attribute of $X$, denoted by $W$, satisfying $H_\infty(P_W) < H_\infty(P_U)$ whose value is also disclosed.
\end{prop}

Proposition~\ref{prop:impossibility_style} is conceptually similar to the impossibility result (specifically, \cite[Thm. 3]{dworkDifficultiesDisclosurePrevention2010a}); yet, it is interpreted differently in our framework: If $U$ is disclosed to provide utility, then $U$ has small entropy and can be estimated accurately using its distribution $P_U$ alone. Since $H_\infty(P_{W}) < H_\infty(P_U)$, then $W$ can too be estimated accurately using its distribution $P_{W}$, even without access to the mechanism. Thus, if disclosing $U$ is not considered as a privacy breach, then disclosing $W$ should not be considered as a privacy breach either. It is also worth mentioning that the proof of Proposition~\ref{prop:impossibility_style} requires no assumptions about the mechanism $P_{Y \mid X}$ other than the fact that it discloses $U$. Hence, the result holds even if we assume that $P_{Y \mid X}$ satisfies $(\epsilon, P_X)$-PML with $\epsilon \geq H_\infty(P_U)$.

As the final topic in this subsection, we discuss \emph{absolute disclosure prevention}, i.e., we investigate conditions ensuring that \emph{no} attribute of $X$ can be disclosed by the mechanism $P_{Y \mid X}$. We show that absolute disclosure prevention can be achieved by mechanisms that have finite leakage capacity (see Definition~\ref{def:leakage_capacity}). Moreover, we prove that these mechanisms guarantee a lower bound on the remaining uncertainty in the value of all (non-constant) deterministic attributes of $X$ for all adversaries in $\mathcal P_\mathcal X$. 

\begin{thm}[Absolute disclosure prevention]
\label{thm:finite_capacity}
If $P_{Y \mid X}$ satisfies $C(P_{Y \mid X}) < \infty$, then for all $P_X \in \mathcal P_\mathcal X$ no attribute of $X$ can be disclosed by $P_{Y \mid X}$. Furthermore, given an arbitrary (non-constant) deterministic function of $X$, denoted by $V$, the remaining uncertainty in the value of $V$ for adversary $Q_X \in \mathcal P_\mathcal X$ is at least 
\begin{equation*}
    H_\infty(Q_{V \mid Y=y}) \geq \log \left (1 + \frac{\min_x q_X(x)}{1 - \min_x q_X(x)} e^{-C(P_{Y\mid X})}\right),
\end{equation*}
for all $y \in \mathcal Y$.
\end{thm}

By Theorem~\ref{thm:pml_ldp}, a mechanism $P_{Y \mid X}$ has finite leakage capacity if and only if it satisfies $(\epsilon, \mathcal P_\mathcal X)$-PML with some finite value of $\epsilon$. As such, the above result contains a similar idea to the no-free-lunch theorem of~\citep{kiferNoFreeLunch2011}. More precisely, \cite[Thm. 2.1]{kiferNoFreeLunch2011} states that it is not possible to discriminate between different instances of the secret $X$ if we guarantee privacy under all possible distributions on the data. That is, utility is essentially destroyed when we make no assumptions about the data-generating distribution. Here, however, we may have a different take on Theorem~\ref{thm:finite_capacity} when viewed through the lens of the local/global dichotomy: Guaranteeing privacy under all possible distributions entails that we no longer can distinguish between local and global features of the data. For example, an attribute $U$ of $X$ may have small entropy under distribution $P_X^{(1)} \in \mathcal P_\mathcal X$ but large entropy under another distribution $P_X^{(2)} \in \mathcal P_\mathcal X$. Since no non-trivial attribute of $X$ can have consistently small entropy under all possible distributions in $\mathcal P_\mathcal X$, then no attribute of $X$ can be considered to capture a property of the whole population. Hence, we inevitably protect all features of $X$. In other words, when we make no assumptions about the data-generating distribution, then a mechanism provides no utility because there is no utility to be provided.     

\subsection{How to Pick \texorpdfstring{$\epsilon$}{epsilon}?}
\label{ssec:what_is_eps}
According to Theorem~\ref{thm:pml_entropy}, if a mechanism $P_{Y \mid X}$ satisfies $(\epsilon, P_X)$-PML, then it will not disclose the value of any attribute of $X$ with entropy larger than $\epsilon$ to any adversary in $\mathcal P_\mathcal X$. Essentially, $\epsilon$ describes where (in terms of entropy) we draw the line between global and local features of $X$, and smaller $\epsilon$ implies stricter privacy requirements. We may select $\epsilon$ by asking: Which features of $X$ do we consider to be sufficiently easy to guess by an analyst who knows $P_X$ such that they may be disclosed without causing a privacy breach? Conversely, we may ask: Which features of $X$ do we wish to keep secret even from an analyst who knows $P_X$ and what is the entropy of those features? In this subsection, we give a few concrete examples of attributes of $X$ that are disclosed at different values of $\epsilon$. We also argue that $\epsilon$ should always remain below the entropy of the data $H_\infty(P_X)$.  

First, we establish the existence of an attribute of $X$ which can be disclosed at the smallest $\epsilon$ compared to all other attributes of $X$. Let $p_\mathrm{min} \coloneqq \min_x p_X(x)$ and $x_\mathrm{min} \in \mathcal X$ be a realization of $X$ with probability $p_\mathrm{min}$. 

\begin{prop}
\label{prop:absolute_disclosure}
Suppose $X$ is distributed according to $P_X$. If the mechanism $P_{Y \mid X}$ satisfies $(\epsilon, P_X)$-PML with $\epsilon < \log \frac{1}{1 -  p_\mathrm{min}}$, then $C(P_{Y \mid X}) < \infty$. Conversely, for each $\epsilon \geq \log \frac{1}{1 -  p_\mathrm{min}}$ there exists an attribute $U$ of $X$ and a mechanism $P_{Y \mid X}$ satisfying $(\epsilon, P_X)$-PML that discloses the value of $U$. 
\end{prop}

The second statement in Proposition~\ref{prop:absolute_disclosure} is proved by constructing an attribute of $X$ which requires the smallest privacy cost (i.e., $\epsilon = \log \frac{1}{1 -  p_\mathrm{min}}$) to be disclosed. This attribute describes a binary random variable that determines whether or not $X$ has value $x \neq x_\mathrm{min}$. Thus, Proposition~\ref{prop:absolute_disclosure} essentially states that giving an affirmative answer (deterministically) to the query \say{Is $X \in \mathcal X \setminus \{x_\mathrm{min}\}$?} induces the smallest privacy cost. Note that an analyst who possesses $P_X$ can correctly predict the answer to this query with probability $1 -  p_\mathrm{min}$ even without access to the mechanism. On the other hand,~\citet[Thm. 1]{issaOperationalApproachInformation2020} construct an attribute of $X$ which takes the largest privacy cost to be disclosed. Roughly speaking, \cite[Thm. 1]{issaOperationalApproachInformation2020} shows that an affirmative answer can be given to the query \say{Is $X \in \{x_\mathrm{min}\}$?} when $\epsilon \geq \log \frac{1}{ p_\mathrm{min}}$. Note that the answer to this query is correctly guessed (without access to the mechanism) with the small probability of $ p_\mathrm{min}$, and that at $\epsilon = \log \frac{1}{ p_\mathrm{min}}$ a mechanism is allowed to answer all possible queries about $X$ error-free.  

Of course, $\epsilon$ should be picked such that no realization of $X$ can be disclosed. That is, $P_{Y \mid X}$ should not be able to deterministically give an affirmative answer to any query of the form \say{Is $X \in \{x\}$?} for any $x \in \mathcal X$. We call this particularly pernicious type of disclosure \emph{singling out}. When $\epsilon < H_\infty(P_X)$, $P_{Y \mid X}$ will not single out the value of $X$. 

\begin{defi}[Singling out]
Suppose $X$ is distributed according to $P_X$. We say that a mechanism $P_{Y \mid X}$ singles out the value of $X$ if $\inf_{y \in \mathcal Y} \; H_\infty(P_{X \mid Y = y}) = 0$.  
\end{defi}

By noting that $X$ is an attribute of $X$, we obtain the following corollary of Theorem~\ref{thm:pml_entropy}. 
\begin{cor}
\label{prop:min_entropy_as_bound}
Suppose $X$ is distributed according to $P_X$. If the mechanism $P_{Y \mid X}$ satisfies $(\epsilon, P_X)$-PML with $\epsilon < H_\infty(P_X)$, then it will not single out the value of $X$.
\end{cor}
Thus, when $P_{Y \mid X}$ has infinite leakage capacity, $H_\infty(P_X)$ must be treated as a strict upper bound on $\epsilon$. In practice, however, $H_\infty(P_X)$ will likely be very large and we should opt for much smaller values of $\epsilon$. We examine this in the example below about a query that could be answered deterministically under favorable conditions. 

\begin{exa}\label{ex:deterministic}
Consider a database $X = (D_1, \ldots, D_n)$ containing $n$ i.i.d entries. Suppose we want to answer the query \say{Are there more than $m$ individuals in the database who identify as female?} as accurately as possible but without disclosing the gender of any individual in the database. When $m \ll n$ or $n-m \ll n$ it may be safe to answer this query deterministically and with no randomness at all. To see why, suppose the individuals in this population identify as female with probability $p \in [0.3,0.7]$. Let $S_i$ be a binary random variable that describes whether or not individual $i \in [n]$ identifies as female, and note that the Markov chain $S_i - D_i - X - Y$ holds. Let $y=1$ denote an affirmative answer to the query and $y=0$ denote a negative answer to the query. 

First, suppose $\frac{m}{n} \leq p$. In this case, answering deterministically with $y=1$ causes the information leakage 
\begin{align*}
    \ell_{P_{XY}}(X \to 1) &= \log\; \frac{\max_{x \in \mathcal X}  p_{Y \mid X=x}(1)}{p_Y(1)} \\
    &= \log\; \frac{1}{1 - P_X\left(\{x : x \text{ contains less than or equal to } m \text{ females} \}\right )}\\
    &= -\log\;  \left(1 - \sum_{k=0}^m \binom{n}{k} \, p^k \cdot (1-p)^{n-k} \right)\\
    &\leq - \log \; \Bigg(1 - \exp \Big(-n D_\mathrm{KL}(\frac{m}{n} \Vert p) \Big) \Bigg),
\end{align*}
where the last inequality follows from a Chernoff bound on the tail of the Binomial distribution~\citep{hagerup1990guided}, and $D_\mathrm{KL}(q \Vert r) = q \log \frac{q}{r} + (1-q) \log \frac{1-q}{1-r}$ denotes the KL-divergence between two Bernoulli distributions with parameters $q,r \in (0,1)$. In Figure~\ref{fig:deterministic_query_ex}, we have plotted the above upper bound on $ \ell_{P_{XY}}(X \to 1)$ for different values of $p$ and $n$. It can be observed that when $\frac{m}{n}$ is small, the amount of information leaked by the deterministic query response is several orders of magnitude smaller than $H_\infty(P_{S_i})$. Note that by Theorem~\ref{thm:pml_entropy}, the gender of no individual will be disclosed by the query response as long as $\ell_{P_{XY}}(X \to 1) < \min\limits_{p \in [0.3, 0.7]} H_\infty(P_{S_i}) = 0.36$. Similarly, when $\frac{m+1}{n} \geq p$, answering the query deterministically with $y=0$ causes the information leakage 
\begin{align*}
    \ell(X \to 0) \leq - \log\; \Bigg(1 - \exp \Big(-n D_\mathrm{KL}(1-\frac{m+1}{n} \Vert 1-p) \Big)\Bigg),\\
\end{align*}
which is very small when $n$ is large and $m$ is close to $n$. 

\begin{figure}
    \centering
    \begin{subfigure}[b]{0.48\textwidth}
         \centering
         \includegraphics[scale=0.18]{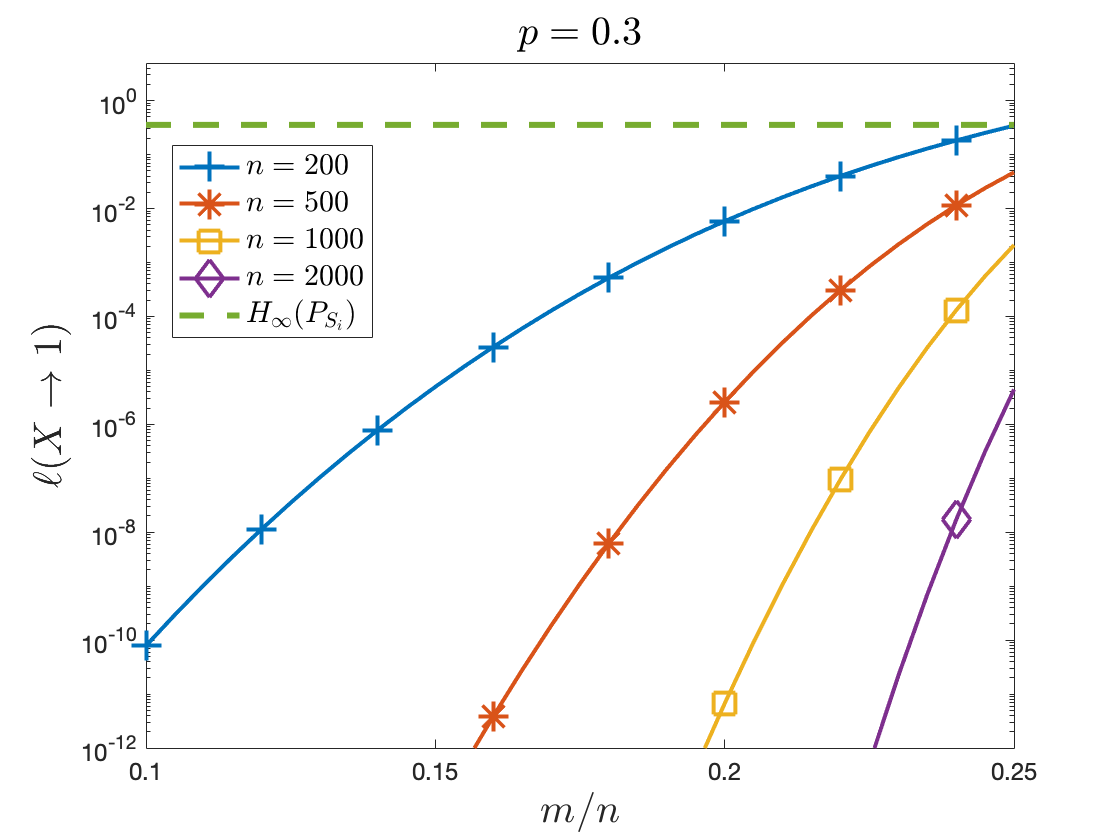}
         \caption{Leakage bounds when $p=0.3$.}
     \end{subfigure}
     \hfill
     \begin{subfigure}[b]{0.48\textwidth}
         \centering
         \includegraphics[scale=0.18]{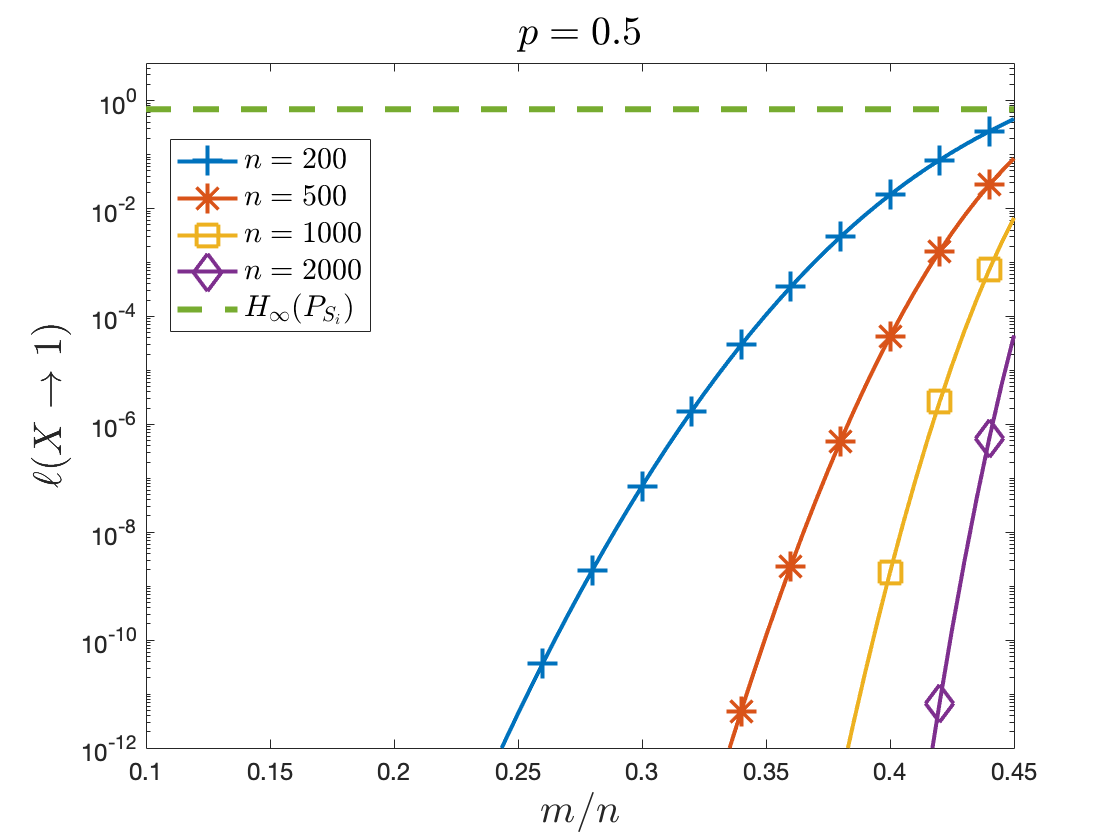}
         \caption{Leakage bounds when $p=0.5$.}
     \end{subfigure}
    \caption{Upper bounds on $\ell_{P_{XY}}(X \to 1)$ in Example~\ref{ex:deterministic} when $p \in \{0.3, 0.5\}$ and $n \in \{200, 500, 1000, 2000\}$.}
    \label{fig:deterministic_query_ex}
\end{figure}
\end{exa}

In conclusion, $\epsilon$ in a PML guarantee is a prior-dependent parameter that is easily interpretable in terms of the entropy of the features of $X$ that we allow to be disclosed. This interpretability is a big advantage over many other privacy definitions, including differential privacy, where no clear guidelines exist that explain how small the privacy parameter should be in order to maintain meaningful privacy guarantees~\citep{dwork2019differential}. 

\section{Inferential Database Privacy}
\label{sec:database_guarantees}
In the previous section, we addressed the main criticisms levied against the inferential perspective on privacy and argued that the inferential view can provide a solid foundation for a consistent privacy framework. In this section, we explore how the inferential view fits in with existing notions of database privacy. Differential privacy, the most widely adopted measure of database privacy, is formulated based on the notion of indistinguishability~\citep{dworkCalibratingNoiseSensitivity}. Roughly speaking, a mechanism is differentially private if all possible outcomes are produced with similar probabilities by two databases that agree on all but one entry. Here, we show that even though differential privacy was proposed as a response to the idea that useful inferential privacy guarantees are impossible~\citep{dworkDifferentialPrivacy}, it is in fact compatible with the inferential perspective through the paradigm of PML. More precisely, we prove that a mechanism satisfies (pure) differential privacy (a) if and only if conditioned on all but one entry, the mechanism releasing information about the remaining entry satisfies $\epsilon$-PML under all possible (product) distributions on $X$, or (b) if and only if the mechanism releasing information about each entry satisfies $\epsilon$-PML under all possible product distributions on $X$. 

Next, we proceed to discuss \emph{free-lunch privacy} and its relationship with PML. \cite{kiferNoFreeLunch2011} argue that differential privacy implicitly assumes that the entries in a database are independent and propose free-lunch privacy as an alternative definition that avoids assumptions about the underlying data-generating distribution. Here, we establish that a mechanism $P_{Y \mid X}$ satisfies free-lunch privacy (a) if and only if $P_{Y \mid X}$ satisfies $\epsilon$-PML under all possible (product) distributions on $X$, or (b) if and only if the mechanism releasing information about each entry satisfies $\epsilon$-PML under all possible distributions on $X$. 

Given that both differential privacy and free-lunch privacy can be expressed as PML constraints, it follows naturally that mechanisms satisfying either of these definitions also guarantee privacy in the sense of PML. To illustrate this point, we examine the counting query and the Laplace mechanism~\citep{dworkCalibratingNoiseSensitivity} through the lens of PML. In doing so, we demonstrate that when the data-generating distribution has large entropy, the privacy cost of using the Laplace mechanism can be considerably smaller than (up to half of) the differential privacy parameter. Thus, PML-based analysis also has the advantage of precisely characterizing the privacy cost by taking into account the data-generating distribution.

Suppose $X$ is a random variable representing a database containing $n$ entries. Given $i \in [n]$, let $D_i$ be the random variable corresponding to the $i$-th entry, which takes values in a finite alphabet $\mathcal D$. Then, each database (realization) $x = (d_1, \ldots, d_n) \in \mathcal D^n$ is an $n$-tuple and $X$ is a sequence of $n$ random variables. Suppose $P_X = P_{D_1, \ldots, D_n}$ denotes the distribution according to which databases are drawn from $\mathcal D^n$. To obtain the probability distribution describing the $i$-th entry we marginalize over the remaining $n-1$ entries, that is, for each $d_i \in \mathcal D$ and $i \in [n]$ we have
\begin{equation*}
    p_{D_i}(d_i) = \sum_{d_{-i} \in \mathcal D^{n-1}} p_{D_i \mid D_{-i} = d_{-i}}(d_i) \; p_{D_{-i}} (d_{-i}),
\end{equation*}
where $d_{-i} \coloneqq (d_1, \ldots, d_{i-1}, d_{i+1}, \ldots, d_n) \in \mathcal D^{n-1}$ is a tuple describing the database with its $i$-th entry removed. Note that this setup is very general in the sense that the entries can be arbitrarily correlated. 

Suppose an analyst poses a query to the database whose answer is returned by the mechanism $P_{Y \mid X}$. Below, we define $\epsilon$-differential privacy\footnote{Technically, Definition~\ref{def:indisting} describes indistinguishability but it is often taken as the definition of differential privacy. This definition is sometimes called \emph{bounded differential privacy} in works such as~\citep{kiferNoFreeLunch2011}.} in our notation. 

\begin{defi}[Differential privacy]
\label{def:indisting}
Given $\epsilon \geq 0$, we say that the mechanism $P_{Y \mid X}$ satisfies $\epsilon$-differential privacy if 
\begin{equation*}
    \sup_{y \in \mathcal Y} \; \max_{\substack{d_i, d_i' \in \mathcal D: \\ i \in [n]}} \; \max_{d_{-i} \in \mathcal D^{n-1}} \log \frac{p_{Y \mid D_i=d_i, D_{-i} = d_{-i}}(y)}{p_{Y \mid D_i=d_i', D_{-i} = d_{-i}}(y)} \leq \epsilon.
\end{equation*}
\end{defi}

Let $\mathcal P_{\mathcal X}$ denote the set of all distributions with full support on $\mathcal X = \mathcal D^n$. Note that assuming the prior distributions on $X$ belong to the set $\mathcal P_{\mathcal X}$ ensures that all conditional probabilities given subsets of the database are well-defined (by making sure that we do not condition on events with probability zero). Furthermore, let $\mathcal Q_\mathcal X$ denote the set of product distributions in $\mathcal P_{\mathcal X}$, that is, $\mathcal Q_\mathcal X \coloneqq \{P_{X} \in \mathcal P_{\mathcal X} \colon P_{X} = \prod_{i=1}^n P_{D_i} \}$. We now show that differential privacy admits multiple different but equivalent formulations in terms of PML. All the results presented in this section are proved in Appendix~\ref{sec:proofs_inferential_database_privacy}. 
\begin{thm}[Differential privacy as a PML constraint]
\label{thm:dp_pml}
Given $\epsilon \geq 0$, the mechanism $P_{Y \mid X}$ satisfies $\epsilon$-differential privacy if and only if 
\begin{enumerate}
    \item $\displaystyle \sup_{y \in \mathcal Y} \sup_{P_X \in \mathcal P_\mathcal X} \; \max_{\substack{d_{-i} \in \mathcal D^{n-1}: \\ i \in [n]}} \; \ell(D_i \to y \mid d_{-i}) \leq \epsilon$, or, 
    \item $\displaystyle \sup_{y \in \mathcal Y} \sup_{P_X \in \mathcal Q_\mathcal X} \; \max_{\substack{d_{-i} \in \mathcal D^{n-1}: \\ i \in [n]}} \; \ell(D_i \to y \mid d_{-i}) \leq \epsilon$, or, 
    \item $\displaystyle \sup_{y \in \mathcal Y} \sup_{P_X \in \mathcal Q_\mathcal X} \; \max_{i \in [n]} \; \ell(D_i \to y) \leq \epsilon$.
\end{enumerate} 
\end{thm}

The first formulation of differential privacy in the above theorem is similar to a result of \cite{dworkCalibratingNoiseSensitivity}. Specifically, \cite[Claim 3]{dworkCalibratingNoiseSensitivity} shows that differential privacy is equivalent to \emph{semantic security}~\cite[Def. 6]{dworkCalibratingNoiseSensitivity}, where semantic security is defined by imposing both an upper bound and a lower bound on the posterior-prior ratio of all binary predicates of the data. The above result can then be considered as a generalization of \cite[Claim 3]{dworkCalibratingNoiseSensitivity} because it only requires an upper bound on the posterior-prior ratio and $D_i$ is not restricted to be binary. 

According to Theorem~\ref{thm:dp_pml}, part (3), differential privacy limits the (non-conditional) information leakage about each entry in the database, provided that the entries are statistically independent. In~\citep{apf24}, it is proved that when the entries in a database are correlated, the PML of a differentially private mechanism can be as large as that of a mechanism that directly releases an entry from the database without any randomization. For completeness, we restate this result below.
 
\begin{thm}[{\citep[Thm. 3]{apf24}}]
\label{thm:dp_correlated_bad}
For each $\delta >0$ and $\epsilon >0$ there exists a correlated database $X = (D_1, \ldots, D_n)$, $i \in [n]$, a mechanism $P_{Y \mid X}$ satisfying $\epsilon$-DP, and $y \in \cY$ such that
\begin{equation*}
    \ell(D_i \to y) > \epsilon_\mathrm{max}(D_i) - \delta, 
\end{equation*}
where 
\begin{equation*}
    \epsilon_\mathrm{max} (D_i) \coloneqq \log \frac{1}{\min\limits_{d \in \cD} \, p_{D_i}(d)}, 
\end{equation*}
is the largest amount of information that any mechanism can leak about $D_i$. 
\end{thm}

Next, we define free-lunch privacy in our notation and show how it can be expressed in terms of PML. In particular, we show that free-lunch privacy limits the (non-conditional) information leakage about each entry in the database, regardless of the distribution.

\begin{defi}[Free-lunch privacy~{\cite[Def. 2.3]{kiferNoFreeLunch2011}}]
\label{def:free_lunch}
Given $\epsilon \geq 0$, we say that the mechanism $P_{Y \mid X}$ satisfies $\epsilon$-free-lunch privacy if 
\begin{equation*}
    \sup_{y \in \mathcal Y} \; \max_{d^n, \tilde d^n \in \mathcal D^n} \log \; \frac{p_{Y \mid X = d^n} (y)}{p_{Y \mid X = \tilde d^n} (y)} \leq \epsilon.
\end{equation*}
\end{defi}

\begin{thm}[Free-lunch privacy as a PML constraint]
\label{thm:free_lunch}
Given $\epsilon \geq 0$, the mechanism $P_{Y \mid X}$ satisfies $\epsilon$-free-lunch privacy if an only if 
\begin{enumerate}
    \item $\displaystyle \sup_{y \in \mathcal Y} \sup_{P_X \in \mathcal P_\mathcal X} \ell(X \to y) \leq \epsilon$, or
    \item $\displaystyle \sup_{y \in \mathcal Y} \sup_{P_X \in \mathcal Q_\mathcal X} \ell(X \to y) \leq \epsilon$, or
    \item $\displaystyle \sup_{y \in \mathcal Y} \sup_{P_X \in \mathcal P_\mathcal X} \; \max_{i \in [n]} \; \ell(D_i \to y) \leq \epsilon$.
\end{enumerate}
\end{thm}

We highlight a few points about the above results. First, note that by the Markov chain $D_i - X - Y$ and the pre-processing inequality for PML~\cite[Lemma 1]{saeidian2023pointwise_it}, $\ell(D_i \to y) \leq \ell(X \to y)$ for all $i \in [n]$, $y \in \mathcal Y$ and $P_X \in \mathcal P_\mathcal X$. Theorem~\ref{thm:free_lunch} then implies that under certain distributions, the amount of information leaking about a single entry can be as large as the information leaking about the whole database. Roughly speaking, this happens when the entropy of the whole dataset is concentrated on a single entry. Second, by comparing (1) in Theorem~\ref{thm:dp_pml}, Theorem~\ref{thm:dp_correlated_bad}, and (3) in Theorem~\ref{thm:free_lunch}, we arrive at a similar conclusion to that of~\cite{kiferNoFreeLunch2011} and \cite{yangBayesianDifferentialPrivacy2015}: that the informed adversary assumption can lead to underestimating the information leaking about the entries in the dataset. Nevertheless, this can happen only when the entries are correlated. Indeed, if we restrict our attention to product distributions, then by (2) and (3) in Theorem~\ref{thm:dp_pml} the worst-case conditional and unconditional leakages become equal. It is also worth mentioning that in the above results, the supremum is never actually attained by any distribution in $\mathcal P_\mathcal X$ or $\mathcal Q_\mathcal X$ (see Remark~\ref{rem:sup_never_attained}). Instead, the proofs construct a sequence of distributions with decreasing (conditional) entropy under which PML converges to the corresponding log-likelihood ratio in the definition of differential privacy or free-lunch privacy. Therefore, when the dataset has large entropy, the amount of information leaking through a mechanism can be noticeably smaller than the $\epsilon$ reported by differential privacy or free-lunch privacy. Below, we use this observation to show how incorporating knowledge about the data-generating distribution into our analysis results in a more accurate privacy risk assessment of the counting query and the Laplace mechanism.  

\subsection{Laplace Mechanism and the Counting Query}
\label{ssec:laplace}
Here, we discuss a concrete example demonstrating that existing mechanisms are compatible with the type of privacy discussed in this paper. We also show that incorporating assumptions about the prior distribution can lead to tighter bounds on the privacy parameter associated with a mechanism. This is because, as discussed earlier, privacy is easier to achieve when the distribution $P_X$ has large entropy compared to when it has small entropy. Note that certain datasets such as financial data for fraud detection or health data for studying rare diseases may naturally contain features with very small entropy. However, in many everyday applications, one encounters high-entropy datasets with more balanced probabilities. In these cases, we can save on the privacy cost paid, and ultimately, achieve more utility. Below, we illustrate this for the archetypical example of a counting query that is answered by the Laplace mechanism~\citep{dworkCalibratingNoiseSensitivity}.  

We consider the third characterization of differential privacy in Theorem~\ref{thm:dp_pml} and restrict the set of product distributions from which $X$ may be drawn. Suppose $X$ is an i.i.d database containing $n$ entries. Consider a predicate $f: \mathcal D \to \{0,1\}$ and suppose we want to answer the counting query \say{What fraction of the entries in the database satisfy $f(d_i)=1$?}. Let $0 \leq c < \frac{1}{2}$ be a constant and assume $P_X \in \mathcal P_c^f$, where
\begin{equation*}
    \mathcal P^f_c = \Big\{P_X \in \mathcal Q_\mathcal X : P_{D_i} (\{d \in \mathcal D : f(d) = 1\}) = p \text{ for all } i\in [n] \text{ and } p \in (c, 1-c) \Big\}.
\end{equation*}
That is, we assume that each entry in the database satisfies the predicate $f$ with probability $p \in (c, 1-c)$. Let $\mathrm{Lap}(\mu, b)$ denote the Laplace distribution with mean $\mu \in \mathbb R$ and scale parameter $b>0$. To answer the counting query, the Laplace mechanism returns an outcome according to the distribution $Y \mid X=(d_1, \ldots, d_n) \sim \mathrm{Lap}(\displaystyle \frac{f(d_1) + \ldots + f(d_n)}{n}, b)$ \citep{dworkCalibratingNoiseSensitivity}.

\begin{prop}\label{prop:laplace}
Consider the predicate $f : \mathcal D \to \{0,1\}$. Suppose $X$ is a database of size $n$ drawn according to a distribution $P_X \in \mathcal P^f_c$. Let $P_{Y \mid X}$ denote the Laplace mechanism with scale parameter $b>0$ answering the counting query corresponding to $f$. Then, the information leaking about each entry in the database is upper bounded by
\begin{equation*}
    \sup_{P_X \in \mathcal P^f_c} \, \sup_{y \in \mathbb R} \, \ell(D_i \to y) \leq \frac{1}{n b} - \log \left((1 - c) + c \exp\left(\frac{1}{n b} \right) \right),
\end{equation*}
for all $i \in [n]$. 
\end{prop}

When $n b$ is large we may use $e^x \geq 1 + x$ and $\log (1+x) \geq x - \frac{x^2}{2}$ for $x \geq 0$ to obtain the simplified bound
\begin{equation*}
    \sup_{P_X \in \mathcal P^f_c} \, \sup_{y \in \mathbb R} \, \ell(D_i \to y) \leq \frac{1 - c}{n b} + \frac{c^2}{2 n^2 b^2},
\end{equation*}
for all $i \in [n]$. Observe that $\frac{1}{nb}$ corresponds to the well-known differential privacy parameter of the Laplace mechanism returning the answer to a query with global sensitivity $\frac{1}{n}$~\citep{dworkCalibratingNoiseSensitivity}. As expected, the above leakage bound also reduces to $\frac{1}{nb}$ when $c=0$, describing the situation where $P_X$ can be any i.i.d distribution in $\mathcal Q_\mathcal X$ with arbitrarily small entropy. On the other hand, when $c$ is close to $\frac{1}{2}$, then the privacy parameter is reduced by almost a factor of $\frac{1}{2}$. Hence, this approach allows us to adjust the privacy cost we pay based on the entropy of the data, and ultimately, achieve higher utility. 

\section{Other Related Works}
\label{sec:related}
\subsection*{Definitional works.} Apart from differential privacy and its extensions (e.g., \citep{dworkOurDataOurselves2006a, mironov2017renyi, dwork2016concentrated, bun2016concentrated,dong2022gaussian}), a large number of privacy definitions have been proposed in the literature, e.g., differential identifiability~\citep{leeDifferentialIdentifiability2012}, membership privacy~\citep{liMembershipPrivacyUnifying2013}, and Pufferfish privacy~\citep{kiferPufferfishFrameworkMathematical2014}. \cite{yangBayesianDifferentialPrivacy2015} introduced \emph{Bayesian differential privacy} which generalizes the informed adversary assumption of differential privacy and considers adversaries who \emph{a priori} know an arbitrary subset of the dataset. \citet{yangBayesianDifferentialPrivacy2015} also show that differential privacy and Bayesian differential privacy are equivalent when the prior is a product distribution. \cite{bassily2013coupled} introduced the framework of \emph{coupled-worlds privacy} which, similarly to PML, is a prior-dependent notion of privacy. Coupled-worlds privacy relaxes differential privacy by requiring that neighboring databases remain indistinguishable under a predefined set of priors instead of all possible priors. A similar definition to coupled-worlds privacy is \emph{noiseless privacy}~\citep{bhaskar2011noiseless} whose goal is to provide noise-free answers to certain queries by leveraging the intrinsic uncertainty in the value of a database as described by a prior distribution. This is a similar idea to what we have considered in our paper as we have shown that some attributes (functions) of the sensitive data $X$ can be disclosed error-free by a mechanism satisfying $\epsilon$-PML, while others will be distorted to avoid revealing too much information about $X$. 

We stress that none of the above notions have as clear an operational meaning as PML. In fact, most definitions have been obtained by formalizing some intuitive understanding of privacy, which then may lead to misunderstandings in what they do or do not guarantee.

\subsection*{\lq Semantics\rq~of differential privacy.}  Several works have interpreted the guarantees of differential privacy or cast it as a constraint in terms of familiar quantities such as total variation distance or mutual information. For example, \citet{kasiviswanathanSemanticsDifferentialPrivacy2015} provided a Bayesian interpretation of differential privacy by showing that the posterior belief of an adversary about the input data does not change much (in terms of total variation distance) whether or not each individual's data is included. \citet{wassermanStatisticalFrameworkDifferential2010a} considered a hypothesis test on the value of a single entry in a database and showed that differential privacy imposes a tradeoff between the Type I and Type II error probabilities. \cite{ghosh2016inferential} defined \emph{inferential privacy} as a constraint on how much an analyst's posterior belief can diverge from her prior belief, and study the inferential privacy guarantees of differentially private mechanisms assuming a certain class of prior distributions. \cite{cuffDifferentialPrivacyMutual2016a} showed that differential privacy is equivalent to a constraint on the conditional mutual information of a mechanism; however, this equivalence is in a weaker sense compared to the one we have established using PML. Finally, \citet{kifer2022bayesian} gave an account of frequentist and Bayesian semantics of multiple variants of differential privacy. Their Bayesian semantics, however, rely solely on posterior-to-posterior comparisons, and posterior-to-prior comparisons are deemed unsuitable due to the results of~\cite{dworkDifficultiesDisclosurePrevention2010a} and \cite{kiferNoFreeLunch2011}. This is exactly the point of view challenged in this paper. 

\section{Conclusions}
In summary, this paper describes a paradigm shift in how privacy is defined that follows from a novel interpretation of the fundamental result of \cite{dworkDifficultiesDisclosurePrevention2010a} about the impossibility of absolute disclosure prevention. According to the definition of privacy presented here, we must distinguish between the properties of the secret $X$ that are predictable using the prior $P_X$ alone and those properties that can be obtained only through the mechanism $P_{Y \mid X}$. This is an important distinction to make especially in applications where the priors are publicly known. For example, it is well-known that a big portion of the human DNA is largely predictable, and intuitively, a mechanism should be able to release this information without it being considered a privacy breach. The advantages of our new paradigm are briefly summarized as follows: 
\begin{itemize}
    \item The view of privacy presented here is inherently Bayesian and inferential. This allows devising privacy-preserving solutions that are adapted to each dataset in terms of entropy, correlations among data points, and so on. 
    \item Privacy guarantees are adapted to the underlying distribution of the data but do not depend on each adversary's perceived information leakage. Essentially, it is possible to distinguish between the useful and necessary information leakage that allows analysts (with inaccurate priors) to learn about the data and the harmful information leakage causing privacy breaches. 
    
    \item The local/global dichotomy fosters a conceptual synergy between privacy and utility, as demonstrated by Example~\ref{ex:good_bad_mean_estimate}. Adopting this perspective, privacy is rendered an actionable goal and is improved by constructing high-quality estimators of the features of the data with suitable convergence properties. 
    
    \item The framework's central privacy notion, PML, is operationally meaningful and precisely defined, with explicit assumptions about adversaries and the general setup.
    \item The privacy parameter in PML guarantees is easily interpretable, providing clear guidelines for parameter selection.
    \item The framework is compatible with and provides insights into existing privacy definitions, including differential privacy.
    \item The framework enables more flexible mechanism designs. Existing mechanisms can be used efficiently to provide meaningful PML-based guarantees and novel mechanisms can be conceived that adjust their randomness to the entropy of the data for increased utility. 
\end{itemize}

Finally, it should also be emphasized that in this paper, we have considered a specific notion of disclosure defined as achieving full certainty about a random variable, or equivalently, zero posterior entropy. Our disclosure prevention result in Theorem~\ref{thm:pml_entropy} relies on this definition to remove dependence on the adversary’s prior knowledge. Nevertheless, alternative definitions of disclosure could lead to different conclusions and may require constraints on the adversary's prior knowledge to ensure disclosure prevention. We leave discussions of other disclosure notions for future work.

\section*{Acknowledgment}
  \noindent This work has been supported by the Swedish Research Council (VR) under the grant 2023-04787 and Digital Futures center within the collaborative project DataLEASH. 

\bibliography{main}
\bibliographystyle{abbrvnat}

\newpage
\appendix
\section{Proofs for Section~\ref{sec:dalenius}}
\label{sec:proofs_impossibility}

\subsection{Proof of Theorem~\ref{thm:ubiq_disclos}}
Consider the Markov chain $U-X-Y$ and suppose $P_{Y \mid X}$ discloses the value of $U$ to adversary $Q_X \in \mathcal P_\mathcal X$. Fix $R_X \in \mathcal P_\mathcal X$. First, we argue that since $Q_X$ and $R_X$ are mutually absolutely continuous, then the posterior distributions $Q_{X \mid Y=y}$ and $R_{X \mid Y=y}$ are also mutually absolutely continuous for all $y \in \mathcal Y$. Let $f(x) = \frac{r_X (x)}{q_X (x)}$ denote the Radon-Nikodym derivative of $R_X$ with respect to $Q_X$ and observe that $f(x) > 0$ for all $x \in \mathcal X$. Fix an arbitrary $x \in \mathcal X$ and $y \in \mathcal Y$. We have 
\begin{align*}
    r_{X \mid Y=y} (x) &= \frac{p_{Y \mid X=x}(y) \cdot  r_X(x)}{r_Y(y)}\\
    &= \frac{p_{Y \mid X=x}(y) \cdot f(x) \cdot  q_X(x)}{r_Y(y)}\\
    &= \frac{q_{X \mid Y=y}(x) \cdot f(x) \cdot  q_Y(y)}{r_Y(y)}\\
    &= q_{X \mid Y=y}(x) \cdot g(x,y),
\end{align*}
where $g(x,y) \coloneqq \frac{f(x) \cdot  q_Y(y)}{r_Y(y)}$ is strictly positive for all $x \in \mathcal X$ and all $y \in \mathcal Y$. Thus, if $q_{X \mid Y=y}(x)$ is positive, then so is $r_{X \mid Y=y}(x)$ and vice versa, proving that the posterior distributions $Q_{X \mid Y=y}$ and $R_{X \mid Y=y}$ are mutually absolutely continuous for all $y \in \mathcal Y$. Next, we note that $g(x,y)$ is bounded above because 
\begin{align*}
    \max_{x \in \mathcal X} \; \sup_{y \in \mathcal Y} \; g(x,y) &= \left( \max_{x \in \mathcal X} f(x) \right) \; \sup_{y \in \mathcal Y} \frac{q_Y(y)}{r_Y(y)}\\
    &= \left( \max_{x \in \mathcal X} f(x) \right) \; \exp \Big(D_\infty (Q_Y \Vert R_Y) \Big)\\
    &\leq \left( \max_{x \in \mathcal X} f(x) \right) \; \exp \Big( D_\infty (Q_X \Vert R_X) \Big),
\end{align*}
where the inequality is due to the data-processing inequality for Rényi divergence~\cite[Thm. 9]{van2014renyi}. The Rényi divergence $D_\infty (Q_X \Vert R_X)$ is also finite since $Q_X$ and $R_X$ are mutually absolutely continuous. Let $c >0$ be a constant satisfying $\max_{x \in \mathcal X} \; \sup_{y \in \mathcal Y} g(x,y) < c$. 

Fix a small $\varepsilon >0$ and an outcome $y \in \mathcal Y$ with $H_\infty (Q_{U \mid Y=y}) < \varepsilon$. Then, there exists $u^* \in \mathcal U$ such that $q_{U \mid Y=y}(u^*) > e^{-\varepsilon}$, which in turn implies that $q_{U \mid Y=y}(u) < 1- e^{-\varepsilon}$ for all $u \neq u^*$. Now, for all $u \neq u^*$ we can write
\begin{align*}
    r_{U \mid Y=y}(u) &= \sum_{x \in \mathcal X} p_{U \mid X=x}(u) \cdot r_{X \mid Y=y}(x) \\
    &= \sum_{x \in \mathcal X} p_{U \mid X=x}(u) \cdot g(x,y) \cdot q_{X \mid Y=y}(x) \\
    &\leq \Big( \max_{x \in \mathcal X} \, g(x,y) \Big) \sum_{x \in \mathcal X} p_{U \mid X=x}(u) \cdot q_{X \mid Y=y}(x) \\
    &= \Big( \max_{x \in \mathcal X} \, g(x,y) \Big) \; q_{U \mid Y=y}(u) \\
    &<  \Big( \max_{x \in \mathcal X} \, g(x,y) \Big) \; (1 - e^{-\varepsilon})\\
    &< c \cdot (1 - e^{-\varepsilon}).
\end{align*}
Thus, we get $r_{U \mid Y=y}(u^*) = 1 - \sum_{u \neq u^*} r_{U \mid Y=y}(u) >  1 - \Big( \abs{\mathcal U} - 1 \Big) \cdot c \cdot (1 - e^{-\varepsilon})$. Finally, taking $\varepsilon \to 0$ yields $r_{U \mid Y=y}(u^*) \to 1$ and we conclude that $\inf_{y \in \mathcal Y} \, H_\infty(R_{U \mid Y=y}) = 0$. In other words, $P_{Y \mid X}$ discloses the value of $U$ to adversary $R_X \in \mathcal P_\mathcal X$.

\subsection{Proof of Theorem~\ref{thm:pml_entropy}}
Fix some $U$ satisfying the Markov chain $U - X - Y$ with entropy $H_\infty(P_U) > \epsilon$, where $P_U = P_{U \mid X} \circ P_X$. First, consider an adversary with prior belief $P_X$. Let $P_{UY} = (P_{U \mid X} \times P_{Y \mid X}) \circ P_{X}$ denote the joint distribution of $U$ and $Y$. Fix an arbitrary $y \in \mathcal Y$. We can write
\begin{subequations}
    \begin{align}
    \ell_{P_{UY}} (U \to y) &= \log \max_{u \in \mathrm{supp}(P_U)} \, \frac{p_{U \mid Y=y}(u)}{p_{U}(u)}\nonumber\\
    &\geq \log \max_{u \in \mathrm{supp}(P_U)} p_{U \mid Y=y}(u) + \log \frac{1}{\max_{u \in \mathrm{supp}(P_U)} p_U(u)}\nonumber\\
    &\geq \log \max_{u \in \mathrm{supp}(P_{U \mid Y=y})} p_{U \mid Y=y}(u) + H_\infty (P_U) \label{subeq:disclosure_1}\\
    &= H_\infty (P_U) - H_\infty(P_{U \mid Y=y}),\nonumber
\end{align}
\end{subequations}
where \eqref{subeq:disclosure_1} is due to the fact that $\mathrm{supp}(P_{U \mid Y=y}) \subseteq \mathrm{supp}(P_U)$ for all $y \in \mathcal Y$. That is, we have 
\begin{align}
\begin{split}
\label{eq:prior_vs_posterior_entropy}
    H_\infty(P_{U \mid Y=y}) &\geq H_\infty (P_U) - \ell_{P_{UY}} (U \to y) \\
    &\geq H_\infty (P_U) - \ell_{P_{XY}} (X \to y),
\end{split}
\end{align}
where the second inequality follows from the pre-processing lemma for PML~\cite[Lemma 1]{saeidian2023pointwise_it}. Now, assuming that $P_{Y \mid X}$ satisfies $(\epsilon, P_X)$-PML, taking the supremum over $y \in \mathcal Y$ yields 
\begin{equation}
\label{eq:no_disclose_true_prior}
    \inf_{y \in \mathcal Y} \, H_\infty(P_{U \mid Y=y}) \geq H_\infty (P_U) - \sup_{y \in \mathcal Y} \, \ell_{P_{XY}} (X \to y) > 0.
\end{equation}
Therefore, $P_{Y \mid X}$ cannot disclose the value of $U$ to adversary $P_X$. Finally, by Theorem~\ref{thm:ubiq_disclos}, $P_{Y \mid X}$ cannot disclose the value of $U$ to any adversary in $\mathcal P_\mathcal X$.

\subsection{Proof of Proposition~\ref{prop:impossibility_style}}
Suppose $U$ is a random variable taking values in the set $\mathcal U = \{1, \ldots, k\}$. Fix a small $\varepsilon >0$ and an outcome $y \in \mathcal Y$ satisfying $H_\infty(P_{U \mid Y=y}) < \varepsilon$. Then, there exists $u \in \mathcal U$ with $p_{U \mid Y=y}(u) > e^{-\varepsilon}$. For simplicity, let this be $u = 1$. We now construct an attribute of $X$ with entropy smaller than $H_\infty(P_U)$ whose value is also disclosed by $P_{Y \mid X}$. Let $W$ be a random variable with alphabet $\mathcal W = \mathcal U$ defined by the conditional pmf
\begin{equation*}
    p_{W \mid U=1} (w) = \begin{cases}
        1, & \mathrm{if} \; w=1, \\
        0, & \mathrm{if} \; w \neq 1,
    \end{cases} 
\end{equation*}
and,
\begin{equation*}
    p_{W \mid U=i} (w) = \begin{cases}
        \lambda , & \mathrm{if} \; w=1, \\
        1 - \lambda, & \mathrm{if} \; w = i,\\
        0, & \mathrm{otherwise,}
    \end{cases} 
    \quad \mathrm{for} \; i=2, \ldots, k,
\end{equation*}
where $0 < \lambda < 1$. Let $P_W = P_{W \mid U} \circ P_U$ and $P_{W \mid Y=y} = P_{W \mid U} \circ P_{U \mid Y=y}$. Observe that $p_W(1) =  \lambda + (1 - \lambda) p_U(1)$, and $p_W(i) = (1 - \lambda) p_U(i)$ for $i = 2, \ldots, k$. Thus, if 
\begin{equation*}
    \lambda > \frac{\max_{u \in [k]} p_U(u) - p_U(1)}{1 - p_U(1)},
\end{equation*}
then $p_W(1) > \max_{u \in [k]} p_U(u)$, which in turn, yields $H_\infty(P_W) < H_\infty(P_U)$. Finally, we have
\begin{align}
\begin{split}
\label{eq:disclosing_W}
    p_{W \mid Y=y}(1) &= \sum_{u \in \mathcal U} p_{W \mid U=u}(1) \, p_{U \mid Y=y}(u)\\
    &\geq p_{W \mid U=1}(1) \, p_{U \mid Y=y}(1)\\
    &> e^{-\varepsilon},
\end{split}
\end{align}
which implies that $H_\infty(P_{W \mid Y=y}) < \varepsilon$. Taking $\varepsilon \to 0$, we conclude that $P_{Y \mid X}$ discloses the value of $W$. 

\subsection{Proof of Theorem~\ref{thm:finite_capacity}}
Consider an adversary $Q_X \in \mathcal P_\mathcal X$ and suppose $C(P_{Y \mid X}) < \infty$. We prove the theorem by contradiction. Suppose $P_{Y\mid X}$ discloses the value of an attribute of $X$, denoted by $U$. Then, for each $\varepsilon >0$ there exists $y \in \mathcal Y$ such that $H_\infty(Q_{U \mid Y=y}) < \varepsilon$, or equivalently, $q_{U \mid Y=y}(u) > e^{-\varepsilon}$ for some $u \in \mathcal U$. Denote this outcome by $u_1$. By Bayes' theorem, we have
\begin{equation*}
    q_{Y \mid U=u_1}(y) = \frac{q_{U \mid Y=y}(u_1) q_Y(y)}{q_U(u_1)} > \frac{q_Y(y)}{q_U(u_1)} \cdot e^{-\varepsilon}. 
\end{equation*}
On the other hand, we also have 
\begin{align*}
    q_{Y \mid U = u_1} (y) = \sum_{x \in \mathcal X} p_{Y \mid X=x}(y) q_{X \mid U = u_1}(x) \leq \max_{x} p_{Y \mid X=x}(y),
\end{align*}
hence, we get $\max_{x} p_{Y \mid X=x}(y) > \frac{q_Y(y)}{q_U(u_1)} \cdot e^{-\varepsilon}$. Furthermore, for all $u \neq u_1$ we have $q_{U \mid Y=y}(u) < 1-e^{-\varepsilon}$. Let $u_2$ be one such outcome. Once again, Bayes' theorem yeilds 
\begin{equation*}
    \sum_{x \in \mathcal X} p_{Y \mid X=x}(y) q_{X \mid U = u_2}(x) = q_{Y \mid U=u_2}(y) = \frac{q_{U \mid Y=y}(u_2) q_Y(y)}{q_U(u_2)} < \frac{q_Y(y)}{q_U(u_2)}(1 - e^{-\varepsilon})
\end{equation*}
which, in turn, implies that $p_{Y \mid X=x}(y) q_{X \mid U = u_2}(x) < \frac{q_Y(y)}{q_U(u_2)}(1 - e^{-\varepsilon})$ for all $x \in \mathcal X$. Now, since $\sum_x q_{X \mid U=u_2}(x) = 1$,  $q_{X \mid U=u_2}(x)$ must be strictly positive for at least one $x \in \mathcal{X}$. Let $x^* \in \mathcal X$ be one such outcome. Hence, we get
\begin{equation*}
    p_{Y \mid X=x^*}(y) < \frac{q_Y(y)}{q_U(u_2) \cdot q_{X \mid U = u_2}(x^*)}(1 - e^{-\varepsilon}). 
\end{equation*}
Finally, we get 
\begin{align*}
    \exp\left(C(P_{Y \mid X})\right) > \frac{\max_x p_{Y \mid X=x}(y)}{p_{Y \mid X=x^*}(y)} > \frac{q_U(u_2) \cdot q_{X \mid U=u_2}(x^*)}{q_U(u_1)} \cdot \frac{e^{-\varepsilon}}{1 - e^{-\varepsilon}} > c \cdot \frac{e^{-\varepsilon}}{1 - e^{-\varepsilon}},
\end{align*}
where $c>0 $ is a suitably small constant. Then, by letting $\varepsilon \to 0$, we conclude that the capacity $C(P_{Y \mid X})$ is infinite which is a contradiction. This proves the first statement.

To prove the second statement, suppose $V$ is a deterministic function of $X$ which is induced by the kernel $P_{V \mid X}$ and takes values in the set $\mathcal V$. Fix an arbitrary $v \in \mathcal V$ and define $\mathcal X_v \coloneqq \{x \in \mathcal X : p_{V \mid X=x} (v) = 1\}$. Note that $p_{V \mid X=x} (v) = 0$ for all $x \notin \mathcal X_v$. Fix an arbitrary $y \in \mathcal Y$ and let $r_\mathrm{min} = \min_x p_{Y \mid X=x}(y)$ and $r_\mathrm{max} = \max_x p_{Y \mid X=x}(y)$. Observe that $\displaystyle \exp(C(P_{Y \mid X})) \geq \frac{r_\mathrm{max}}{r_\mathrm{min}}$. Let $Q_{VY} = (P_{V \mid X} \times P_{Y \mid X}) \circ Q_X$ denote the joint distribution of $V$ and $Y$. We can write 
\begin{align*}
    q_{V \mid Y=y} (v) = \frac{q_{V Y} (v,y)}{q_Y(y)} &=\frac{\sum_{x \in \mathcal X} p_{V \mid X=x}(v) \, p_{Y \mid X=x}(y) \, q_X(x)}{\sum_{x \in \mathcal X} p_{Y \mid X=x}(y) \, q_X(x)}\\[0.5em]
    &= \frac{\sum_{x \in \mathcal X_v} p_{Y \mid X=x}(y) q_X(x)}{\sum_{x \in \mathcal X} p_{Y \mid X=x}(y) q_X(x)}\\[0.5em]
    &= \frac{1}{\displaystyle 1 + \frac{\sum_{x \notin \mathcal X_v} p_{Y \mid X=x}(y) q_X(x)}{\sum_{x \in \mathcal X_v} p_{Y \mid X=x}(y) q_X(x)}}\\
    & \leq \frac{1}{\displaystyle 1 + \frac{r_\mathrm{min} \; \left( 1 - Q_X(\mathcal X_v) \right)}{r_\mathrm{max} \; Q_X(\mathcal X_v)}}\\[0.5em]
    &\leq \frac{1}{\displaystyle 1 + \frac{\min_x q_X(x)}{\exp(C(P_{Y \mid X})) (1 - \min_x q_X(x))}}. 
\end{align*}

Thus, we get 
\begin{align*}
    H_\infty (Q_{V \mid Y=y}(v)) &= \log \frac{1}{\max_v q_{V \mid Y=y}(v)}\\
    &\geq \log \left( 1 + \frac{\min_x q_X(x)}{1 - \min_x q_X(x)} e^{-C(P_{Y \mid X})} \right). 
\end{align*}

\subsection{Proof of Proposition~\ref{prop:absolute_disclosure}}
\label{subsec:proof_absolute_disclosure}
Suppose $C(P_{Y \mid X}) = \infty$. Then, for each $\varepsilon >0$, there exists $y_\varepsilon \in \mathcal Y$ such that 
\begin{equation*}
    \frac{\max_x p_{Y \mid X=x}(y_\varepsilon)}{\min_x p_{Y \mid X=x}(y_\varepsilon)} \geq \frac{1}{\varepsilon}.
\end{equation*}
Let $\bar x_\varepsilon \in \argmax_x p_{Y \mid X=x}(y_\varepsilon)$ and $\ubar x_\varepsilon \in \argmin_x p_{Y \mid X=x}(y_\varepsilon)$. We have 
\begin{align*}
    \sup_{y \in \mathcal Y} \; \ell_{P_{XY}}(X \to y) &= \sup_{y \in \mathcal Y}  \; \log \frac{\max_{x \in \mathcal X} p_{Y \mid X=x}(y)}{p_{Y}(y)}\nonumber\\
    &\geq \sup_{\varepsilon > 0} \; \log \frac{p_{Y \mid X=\bar x_\varepsilon}(y_\varepsilon)}{p_{Y \mid X=\ubar x_\varepsilon}(y_\varepsilon) \, p_X(\ubar x_\varepsilon) + \sum_{x \neq \ubar x_\varepsilon} p_{Y \mid X= x}(y_\varepsilon) \, p_{X}(x)}\nonumber\\
    &\geq \sup_{\varepsilon > 0} \; \log \frac{p_{Y \mid X=\bar x_\varepsilon}(y_\varepsilon)}{\varepsilon \, p_{Y \mid X=\bar x_\varepsilon}(y_\varepsilon) \, p_X(\ubar x_\varepsilon) + \sum_{x \neq \ubar x_\varepsilon} p_{Y \mid X= \bar x_\varepsilon}(y_\varepsilon) \, p_{X}(x)}\nonumber\\
    &= \sup_{\varepsilon > 0} \; \log \frac{1}{\varepsilon \, \, p_X(\ubar x_\varepsilon) + \sum_{x \neq \ubar x_\varepsilon} p_{X}(x)}\nonumber\\
    &\geq \log \, \frac{1}{1 - p_{\mathrm{min}}}.
\end{align*}
Therefore, no mechanism with infinite leakage capacity can satisfy $(\epsilon,P_X)$-PML with $\epsilon < \log \frac{1}{1-p_{\mathrm{min}}}$.

To prove the second part of the statement, it suffices to construct a mechanism $P_{Y \mid X}$ satisfying $\log \frac{1}{1 - p_\mathrm{min}}$-PML, and an attribute $U$ of $X$ which is disclosed by an outcome of $P_{Y \mid X}$. Consider the binary random variable $U(X) = \mathds 1_{\mathcal X \setminus \{x_\mathrm{min}\}} (X)$ which is a deterministic function of $X$. 

The posterior distribution $P_{X \mid U}$ is given by
\begin{equation*}
    p_{X \mid U=0}(x) = \begin{cases} 
    1, & \mathrm{if} \; x=x_\mathrm{min},\\
    0, & \mathrm{if} \; x \neq x_\mathrm{min},
    \end{cases}
\end{equation*}
and
\begin{equation*}
    p_{X \mid U=1}(x) = \begin{cases} 
    0, & \mathrm{if} \; x=x_\mathrm{min},\\
    \frac{P_X(x)}{1 - p_\mathrm{min}}, & \mathrm{if} \; x \neq x_\mathrm{min}.
    \end{cases}
\end{equation*}
Let $\alpha > 0$ be a small constant. Suppose $Y$ be a binary random variable induced by the mechanism $P_{Y \mid X}$ defined as
\begin{equation*}
    p_{Y \mid X=x}(0) = \begin{cases}
        0, & \mathrm{if} \; x=x_\mathrm{min},\\
        \alpha, & \mathrm{if} \; x \neq x_\mathrm{min},
    \end{cases}
\end{equation*}
and,
\begin{equation*}
    p_{Y \mid X=x}(1) = \begin{cases} 
    1, & \mathrm{if} \; x=x_\mathrm{min},\\
    1 - \alpha, & \mathrm{if} \; x \neq x_\mathrm{min}.
    \end{cases}
\end{equation*}
Then, we have 
\begin{gather*}
    \ell(X \to 0) = \log \frac{1}{1 - p_\mathrm{min}},\\  
    \ell(X \to 1) = \log \frac{1}{1 - \alpha (1- p_\mathrm{min})}.
\end{gather*}
Note that for small enough $\alpha$ we have $\ell(X \to 0) > \ell(X \to 1)$. Hence, $P_{Y \mid X}$ satisfies $\log \frac{1}{1 - p_\mathrm{min}}$-PML. We now verify that $P_{Y \mid X}$ discloses the value of $U$. Let $P_{Y\mid U} = P_{Y \mid X} \circ P_{X \mid U}$. We have
\begin{equation*}
    p_{Y \mid U=u}(0) = \sum_{x} p_{Y \mid X=x}(0) p_{X \mid U=u}(x) = \begin{cases}
        0, & \mathrm{if} \; u=0,\\
        \alpha, & \mathrm{if} \; u=1.\\
    \end{cases}
\end{equation*}
That is, if the adversary observes $y=0$ she will be certain that $U$ has value $u = 1$. Hence, the mechanism $P_{Y \mid X}$ discloses the value of $U$, which completes the proof.

\section{Proofs for Section~\ref{sec:database_guarantees}}
\label{sec:proofs_inferential_database_privacy}
\subsection{Proof of Theorem~\ref{thm:dp_pml}}
Fix an arbitrary $i \in [n]$, $y \in \mathcal Y$, and $P_X \in \mathcal P_\mathcal X$. We have
\begin{subequations}
\begin{align}
    &\max_{d_{-i} \in \mathcal D^{n-1}} \; \exp \Big( \ell(D_i \to y \mid d_{-i}) \Big) \nonumber\\
    &= \max_{d_{-i} \in \mathcal D^{n-1}} \; \max_{d_i \in \mathrm{supp}(P_{D_i \mid D_{-i} = d_{-i}})} \frac{p_{Y \mid D_{-i} = d_{-i}, D_i=d_i} (y)}{p_{Y \mid D_{-i} = d_{-i}} (y)} \label{subeq:dp_as_pml_0}\\
    &= \max_{d_{-i} \in \mathcal D^{n-1}} \; \max_{d_i \in \mathcal D} \frac{p_{Y \mid D_{-i} = d_{-i}, D_i=d_i} (y)}{p_{Y \mid D_{-i} = d_{-i}} (y)} \label{subeq:dp_as_pml_1}\\
    &= \max_{d_{-i} \in \mathcal D^{n-1}} \max_{d_i \in \mathcal D} \frac{p_{Y \mid D_{-i} = d_{-i}, D_i=d_i} (y)}{\sum\limits_{d_i' \in \mathcal D} p_{Y \mid D_{-i} = d_{-i} , D_i =d_i'} (y) \; p_{D_i \mid D_{-i} = d_{-i}} (d_i')} \nonumber\\
    &\leq \max_{d_{-i} \in \mathcal D^{n-1}} \max_{d_i \in \mathcal D} \; \frac{p_{Y \mid D_{-i} = d_{-i}, D_i=d_i} (y)}{\Big(\min\limits_{d_i' \in \mathcal D} p_{Y \mid D_{-i} = d_{-i} , D_i =d_i'} (y)\Big)  \sum\limits_{d_i' \in \mathcal D} p_{D_i \mid D_{-i}=d_{-i}}(d_i')} \label{subeq:dp_as_pml_2}\\
    &= \max_{d_{-i} \in \mathcal D^{n-1}} \max_{d_i \in \mathcal D} \; \frac{p_{Y \mid D_{-i} = d_{-i}, D_i=d_i} (y)}{\min\limits_{d_i' \in \mathcal D}  \, p_{Y \mid D_{-i} = d_{-i} , D_i =d_i'} (y)} \nonumber\\
    &= \max_{d_{-i} \in \mathcal D^{n-1}} \max_{d_i,d_i' \in \mathcal D} \; \frac{p_{Y \mid D_{-i} = d_{-i}, D_i=d_i} (y)}{p_{Y \mid D_{-i} = d_{-i} , D_i =d_i'} (y)},\nonumber
\end{align}
\end{subequations}
where \eqref{subeq:dp_as_pml_0} follows from Definition~\ref{def:cond_pml}, and \eqref{subeq:dp_as_pml_1} uses the fact that $\mathrm{supp}(P_{D_i \mid D_{-i} = d_{-i}}) = \mathcal D$ for each $P_X \in \mathcal P_\mathcal X$.

Next, we show that the above inequality holds with equality for a product distribution $P_X^* \in \mathcal Q_\mathcal X$. This then proves that (1) and (2) in the statment of the theorem are equivalent to each other and to differential privacy. Let $\varepsilon > 0$ be a small constant. Suppose $P_X^* = \Pi_{i=1}^n P_{D_i}^*$, where 
\begin{align*}
    p^*_{D_i} (d_i') \coloneqq \begin{cases}
        1 - \varepsilon, &\text{for some } d_i' \in \argmin\limits_{\tilde{d_i} \in \mathcal D} \,  p_{Y \mid D_{-i} = d_{-i} , D_i = \tilde d_i} (y),\\
        \frac{\varepsilon}{\abs{\mathcal D} - 1}, & \text{otherwise}.
    \end{cases}
\end{align*} 
Then, $\sum\limits_{d_i' \in \mathcal D} p_{Y \mid D_{-i} = d_{-i} , D_i =d_i'} (y) \; p^*_{D_i} (d_i') \to \min\limits_{d_i' \in \mathcal D}  \, p_{Y \mid D_{-i} = d_{-i} , D_i =d_i'} (y)$ as $\varepsilon \to 0$. Thus, inequality~\eqref{subeq:dp_as_pml_2} holds with equality for $P_X^*$. 

Now, we show that (3) in the statement of the theorem is also equivalent to differential privacy. Fix an arbitrary $i \in [n]$ and $y \in \mathcal Y$. Note that each $P_X \in \mathcal Q_\mathcal X$ can be written as $P_X = P_{D_i} \times P_{D_{-i}}$; hence, we can optimize over $P_{D_i}$ and $P_{D_{-i}}$ separately: 
\begin{subequations}
\begin{align}
    \sup_{P_{D_{-i}}} \; \sup_{P_{D_i}} \; \exp \Big(\ell(D_i \to y)\Big) &= \sup_{P_{D_{-i}}} \; \max_{d_i, d_i^\prime} \; \frac{p_{Y \mid D_i = d_i} (y)}{p_{Y \mid D_i = d_i^\prime} (y)} \label{subeq:product_0}\\ 
    &= \max_{d_i, d_i^\prime} \; \sup_{P_{D_{-i}}} \; \frac{\sum_{d_{-i} } p_{Y \mid D_i = d_i, D_{-i} = d_{-i}} (y) \; p_{D_{-i}} (d_{-i})}{\sum_{d_{-i}} p_{Y \mid D_i = d_i^\prime, D_{-i} = d_{-i}} (y) \; p_{D_{-i}} (d_{-i})} \nonumber\\ 
    &\leq \max_{d_i, d_i^\prime} \; \max_{d_{-i}} \frac{p_{Y \mid D_i = d_i, D_{-i} = d_{-i}} (y)}{p_{Y \mid D_i = d_i^\prime, D_{-i} = d_{-i}} (y)} \label{subeq:product_2},
\end{align}
\end{subequations}
where~\eqref{subeq:product_0} is due to Theorem~\ref{thm:pml_ldp}. To show that inequality~\eqref{subeq:product_2} can be attained, for fixed $d_i$ and $d_i'$ let 
\begin{equation*}
    d_{-i}^* = (d_1^*, \ldots, d_{i-1}^*, d_{i+1}^*, \ldots, d_n^*) \in \argmax_{\tilde d_{-i}} \frac{p_{Y \mid D_i = d_i, D_{-i} = \tilde d_{-i}} (y)}{p_{Y \mid D_i = d_i^\prime, D_{-i} =\tilde d_{-i}} (y)}.
\end{equation*}
Consider the pmf $q^*_{D_j}$ defined by 
\begin{align}
\label{eq:good_product}
    q^*_{D_j} (d_j) \coloneqq \begin{cases}
        1 - \varepsilon, & d_j = d_j^*,\\
        \frac{\varepsilon}{\abs{\mathcal D} - 1}, & \text{otherwise},
    \end{cases}
\end{align}
for $j \neq i$. Let $q_{D_{-i}}^* = \prod_{j \neq i} q_{D_j}^*$ which satisfies $q_{D_{-i}}^* (d_{-i}^*) = (1 - \varepsilon)^{n-1}$, and $q_{D_{-i}}^* (d_{-i}) \leq \frac{\varepsilon}{\abs{\mathcal D} - 1} (1 - \varepsilon)^{n-2}$ for all $d_{-i} \neq d_{-i}^*$. Then, for fixed $n$, 
\begin{equation*}
    \frac{\sum_{d_{-i} } p_{Y \mid D_i = d_i, D_{-i} = d_{-i}} (y) \; q^*_{D_{-i}} (d_{-i})}{\sum_{d_{-i}} p_{Y \mid D_i = d_i^\prime, D_{-i} = d_{-i}} (y) \; q^*_{D_{-i}} (d_{-i})} \to \max_{d_{-i}} \frac{p_{Y \mid D_i = d_i, D_{-i} = d_{-i}} (y)}{p_{Y \mid D_i = d_i^\prime, D_{-i} = d_{-i}} (y)},
\end{equation*}
as $\varepsilon \to 0$. Thus, inequality~\eqref{subeq:product_2} holds with equality for distribution $Q_{D_{-i}}^*$, which completes the proof.\hfill $\square$

\begin{rem}
\label{rem:sup_never_attained}
It is important to note that in all of the formulations above the supremum is never actually attained by any distribution in $\mathcal P_\mathcal X$ or $\mathcal Q_\mathcal X$. For example, consider statement (1). Fix $i \in [n]$ and $d_{-i} \in \mathcal D^{n-1}$, and suppose there exists $Q^*_{D_i \mid D_{-i} = d_{-i}}$ such that 
\begin{align}
\begin{split}
\label{eq:sup_never_attained}
    \sum_{d_i' \in \mathcal D} p_{Y \mid D_{-i} = d_{-i} , D_i =d_i'} (y) \; q^*_{D_i \mid D_{-i} = d_{-i}} (d_i')= \min_{\tilde{d_i} \in \mathcal D} p_{Y \mid D_{-i} = d_{-i} , D_i = \tilde{d_i}} (y).
\end{split}
\end{align}
This equality holds if and only if $p_{Y \mid D_{-i} = d_{-i} , D_i =d_i'} (y) = \min\limits_{\tilde{d_i} \in \mathcal D} p_{Y \mid D_{-i} = d_{-i} , D_i = \tilde{d_i}} (y)$ for all $d_i' \in \mathcal D$. Since \eqref{eq:sup_never_attained} must hold for all $i$ and all $d_{-i}$, $p_{Y \mid D_{-i} = d_{-i} , D_i =d_i} (y)$ must be a constant that does not depend on $d_i$ and $d_{-i}$ for all $y$. However, this implies that $X$ and $Y$ are independent. 
\end{rem}

\subsection{Proof of Theorem~\ref{thm:free_lunch}}
The proof is fairly similar to the proof of Theorem~\ref{thm:dp_pml}; thus, some details are removed. First, note that it follows directly from Theorem~\ref{thm:pml_ldp} that (1) in the statement of the theorem is equivalent to $\epsilon$-free-lunch privacy. To prove that (2) is equivalent to (1) we show that $\sup_{P_X \in \mathcal Q_\mathcal X} \ell(X \to y) \geq \sup_{P_X \in \mathcal P_\mathcal X} \ell(X \to y)$ for all $y \in \mathcal Y$ since the reverse inequality holds trivially. Consider the database $x^* = (d_1^*, \ldots, d_n^*) \in \argmin_x P_{Y \mid X=x}(y).$ We can use a construction similar to~\eqref{eq:good_product} to obtain a product distribution $Q^*_X$ that satisfies $q^*_X(x^*) = (1 - \varepsilon)^n$ while $q^*_X(x) \leq \frac{\varepsilon}{\abs{\mathcal D} - 1} (1 - \varepsilon)^{n-1}$ for all $x \neq x^*$. Then, we get
\begin{align*}
    \sup_{P_X \in \mathcal Q_\mathcal X} \exp \Big( \ell(X \to y) \Big) &\geq \exp \Big(\ell_{P_{Y \mid X} \times Q^*_X}(X \to y) \Big)\\
    &= \frac{\max_x p_{Y \mid X=x}(x)}{\sum_{x'} p_{Y \mid X=x'}(y) q^*_X(x)}\\
    &\geq \frac{\max_x p_{Y \mid X=x}(x)}{(1 - \varepsilon)^n p_{Y \mid X=x^*}(y) + \frac{\varepsilon}{\abs{\mathcal D} - 1} (1 - \varepsilon)^{n-1} \sum_{x' \neq x^*} p_{Y \mid X=x'}(y)}. \\
\end{align*}
For fixed $n$, letting $\varepsilon \to 0$ yields 
\begin{align*}
    \sup_{P_X \in \mathcal Q_\mathcal X} \exp \Big(\ell(X \to y) \Big) &\geq  \frac{\max_x p_{Y \mid X=x}(x)}{\min_{x'} p_{Y \mid X=x'}(y)} \\
    &= \sup_{P_X \in \mathcal P_\mathcal X} \exp \Big(\ell(X \to y) \Big),
\end{align*}
as desired. 

Finally, we show that (3) is equivalent to (1). By the pre-processing inequality for PML~\cite[Lemma 1]{saeidian2023pointwise_it} we have $\ell(D_i \to y) \leq \ell(X \to y)$ for all $i \in [n]$, $y \in \mathcal Y$, and $P_X \in \mathcal P_\mathcal X$. So, we show that  $\sup_{P_X \in \mathcal P_\mathcal X} \max_{i \in [n]} \ell(D_i \to y) \geq \sup_{P_X \in \mathcal P_\mathcal X} \ell(X \to y)$ for all $y \in \mathcal Y$. Fix an arbitrary $i \in [n]$. We write $P_X = P_{D_i} \times P_{D_{-i} \mid D_i}$ and optimize over $P_{D_i}$ and $P_{D_{-i} \mid D_i}$ separately:
\begin{subequations}
\begin{align}
    \sup_{P_{D_{-i} \mid D_i}} \; \sup_{P_{D_i}} \; \exp \Big(\ell(D_i \to y)\Big) &=\; \sup_{P_{D_{-i} \mid D_i}} \; \max_{d_i} \; \sup_{P_{D_i}}\; \frac{p_{Y \mid D_i = d_i} (y)}{p_{Y}(y)} \nonumber\\ 
    &= \sup_{P_{D_{-i} \mid D_i}} \; \max_{d_i, d_i^\prime} \; \frac{p_{Y \mid D_i = d_i} (y)}{p_{Y \mid D_i = d_i^\prime} (y)} \label{subeq:indiv_1}\\ 
    &= \max_{d_i, d_i^\prime} \sup_{P_{D_{-i} \mid D_i}} \; \frac{\sum_{d_{-i} } p_{Y \mid D_i = d_i, D_{-i} = d_{-i}} (y) \; p_{D_{-i} \mid D_i = d_i} (d_{-i})}{\sum_{d_{-i}^\prime } p_{Y \mid D_i = d_i^\prime, D_{-i} = d_{-i}^\prime} (y) \; p_{D_{-i} \mid D_i = d_i^\prime} (d_{-i}^\prime)} \nonumber,
\end{align}
\end{subequations}
where~\eqref{subeq:indiv_1} follows from Theorem~\ref{thm:pml_ldp}. 

Consider the kernel $P^*_{D_{-i} \mid D_i}$ described by
\begin{align*}
    p^*_{D_{-i} \mid D_{i} = d_i} (d_{-i}) \coloneqq\begin{cases}
        1 - \varepsilon, & \text{for some } d_{-i} \in \argmax\limits_{{\tilde d_{-i}}} p_{Y \mid D_{-i} = \tilde d_{-i} , D_i =d_i} (y),\\
        \frac{\varepsilon}{\abs{\mathcal D}^{n-1} - 1}, & \text{otherwise},
    \end{cases}
\end{align*}
and 
\begin{align*}
    p^*_{D_{-i} \mid D_{i} = d_i^\prime} (d_{-i}) \coloneqq \begin{cases}
        1 - \varepsilon, & \text{for some } d_{-i} \in \argmin\limits_{\tilde d_{-i}} p_{Y \mid D_{-i} = \tilde d_{-i} , D_i =d_i^\prime} (y),\\
        \frac{\varepsilon}{\abs{\mathcal D}^{n-1} - 1}, & \text{otherwise}.
    \end{cases}
\end{align*}
Then, we get 
\begin{subequations}
\begin{align}
    &\sup_{P_{D_{-i} \mid D_i}} \; \sup_{P_{D_i}} \; \exp \Big(\ell(D_i \to y)\Big) \nonumber\\
    &\geq \max_{d_i, d_i^\prime}  \; \frac{\sum_{d_{-i} } p_{Y \mid D_i = d_i, D_{-i} = d_{-i}} (y) \; p^*_{D_{-i} \mid D_i = d_i} (d_{-i})}{\sum_{d_{-i}^\prime } p_{Y \mid D_i = d_i^\prime, D_{-i} = d_{-i}^\prime} (y) \; p^*_{D_{-i} \mid D_i = d_i^\prime} (d_{-i}^\prime)} \nonumber\\ 
    &= \max_{d_i, d_i^\prime} \; \frac{\max_{d_{-i}} p_{Y \mid D_i = d_i, D_{-i} = d_{-i}} (y)}{\min_{d_{-i}^\prime} p_{Y \mid D_i = d_i^\prime, D_{-i} = d_{-i}^\prime} (y)} \label{subeq:indiv_2}\\ 
    &= \max_{d_i, d_i^\prime} \; \max_{d_{-i}, d_{-i}^\prime} \; \frac{p_{Y \mid D_i = d_i, D_{-i} = d_{-i}} (y)}{p_{Y \mid D_i = d_i^\prime, D_{-i} = d_{-i}^\prime} (y)}, \nonumber
\end{align}
\end{subequations}
where \eqref{subeq:indiv_2} follows by letting $\varepsilon \to 0$.

\subsection{Proof of Proposition~\ref{prop:laplace}}
Given $i \in [n]$, let $B_i = f(D_i)$ be a binary random variable that determines whether or not entry $D_i$ satisfies the predicate $f$. Since the outcome of the Laplace mechanism depends on $D_i$ only through $B_i$ the Markov chain $D_i - B_i - Y$ holds and $\ell(D_i \to y) \leq \ell(B_i \to y)$ for all outcomes $y \in \mathcal Y$. Thus, we may without loss of generality assume that $D_i = B_i$, that is, we assume that the database is binary. 

For notational simplicity suppose $i=1$. We write 
\begin{align*}
    \sup_{y \in \mathcal Y} \; \ell(D_1 \to y) &= \sup_{y \in \mathcal Y} \; \log \; \frac{\max\limits_{d_1 \in \{0,1\}}p_{Y \mid D_1=d_1}(y)}{p_Y(y)}\\
    &= \sup_{y \in \mathcal Y} \; \log \; \frac{\max\limits_{d_1 \in \{0,1\}} \mathbb E_{D_{-1}}\Big[\displaystyle \exp(-\frac{\abs{y - \frac{d_1}{n} - \frac{S_{-1}}{n}} }{b})\Big]}{\mathbb E_X\Big[\displaystyle \exp(-\frac{\abs{y - \frac{S_X}{n}} }{b})\Big]},
\end{align*}
where $S_{-1} \coloneqq \sum_{i=2}^n D_i$ and $S_X \coloneqq \sum_{i=1}^n D_i$. We argue that it is sufficient to consider $y>1$ and $y<0$. This is because in the numerator we have 
\begin{multline*}
    \mathbb E_{D_{-1}}\Big[\displaystyle \exp(-\frac{\abs{y - \frac{d_1}{n} - \frac{S_{-1}}{n}} }{b})\Big] \\
    \leq \min \left\{\mathbb E_{D_{-1}}\Big[\displaystyle \exp(-\frac{y - \frac{d_1}{n} - \frac{S_{-1}}{n} }{b})\Big], \mathbb E_{D_{-1}}\Big[\displaystyle \exp(\frac{y - \frac{d_1}{n} - \frac{S_{-1}}{n} }{b})\Big] \right\}.
\end{multline*}
Furthermore, the mapping $y \mapsto \mathbb E_X\Big[\displaystyle \exp(-\frac{\abs{y - \frac{S_X}{n}} }{b})\Big]$ in the denominator is increasing in $(-\infty, p]$ and decreasing in $[p, \infty)$ since $S_X$ is a Binomial random variable with success probability $p$.  

Now, if $y>1$, then 
\begin{align*}
    \ell(D_1 \to y) &= \log \; \frac{\max\limits_{d_1 \in \{0,1\}} \mathbb E_{D_{-1}}\Big[\displaystyle \exp(-\frac{y - \frac{d_1}{n} - \frac{S_{-1}}{n}}{b})\Big]}{\mathbb E_X\Big[\displaystyle \exp(-\frac{y - \frac{S_X}{n}}{b})\Big]}\\
    &= \log \; \frac{\max\limits_{d_1 \in \{0,1\}} \mathbb E_{D_{-1}}\Big[\displaystyle \exp(\frac{d_1}{n b} + \frac{S_{-1}}{n b})\Big]}{\mathbb E_X\Big[\displaystyle \exp(\frac{S_X}{n b})\Big]}\\
    &= \frac{1}{n b} + \log \; \frac{\mathbb E_{D_{-1}} [\displaystyle \exp(\frac{D_2 + \ldots + D_n}{n b})]}{\mathbb E_X [\displaystyle\exp(\frac{D_1 + \ldots + D_n}{n b})]}\\
    &= \frac{1}{n b} + \log \; \frac{\Pi_{j=2}^n  \mathbb E[\displaystyle \exp(\frac{D_j}{n b})]}{\Pi_{j=1}^n  \mathbb E[ \displaystyle\exp(\frac{D_j}{n b})]}\\
    &= \frac{1}{n b} - \log \; \left( (1 - p) + p \exp\left(\frac{1}{n b} \right) \right)\\
    &\leq \frac{1}{n b} - \log \; \left( (1 - c) + c \exp\left(\frac{1}{n b} \right) \right),
\end{align*}
where the inequality is due to the fact that the mapping $p \mapsto (1-p) + p \exp(\frac{1}{nb})$ is increasing in $p$. Similarly, if $y<0$, then  
\begin{align*}
    \ell(D_1 \to y) &= \log \; \frac{\max\limits_{d_1 \in \{0,1\}} \mathbb E_{D_{-1}}\Big[\displaystyle \exp(\frac{y - \frac{d_1}{n} - \frac{S_{-1}}{n}}{b})\Big]}{\mathbb E_X\Big[\displaystyle \exp(\frac{y - \frac{S_X}{n}}{b})\Big]}\\
    &= \log \; \frac{\max\limits_{d_1 \in \{0,1\}} \mathbb E_{D_{-1}}\Big[\displaystyle \exp(-\frac{d_1}{n b} - \frac{S_{-1}}{n b})\Big]}{\mathbb E_X\Big[\displaystyle \exp(-\frac{S_X}{n b})\Big]}\\
    &= \log \; \frac{\mathbb E_{D_{-1}} [\displaystyle \exp(-\frac{D_2 + \cdots + D_n}{n b})]}{\mathbb E_X [\displaystyle \exp(-\frac{D_1 + \ldots + D_n}{n b})]}\\
    &= \log \; \frac{\Pi_{j=2}^n  \mathbb E[\displaystyle \exp(-\frac{D_j}{n b})]}{\Pi_{j=1}^n  \mathbb E[\displaystyle \exp(-\frac{D_j}{n b})]}\\
    &= \frac{1}{n b} - \log \; \left(p + (1-p) \exp\left(\frac{1}{n b} \right) \right)\\
    &\leq \frac{1}{n b} - \log \; \left( (1 - c) + c \exp\left(\frac{1}{n b} \right) \right),
\end{align*}
where the last inequality is due to the fact that the mapping $p \mapsto p + (1-p) \exp(\frac{1}{nb})$ is decreasing in $p$. We conclude that 
\begin{equation*}
    \sup_{P_X \in \mathcal P^f_c} \, \sup_{y \in \mathbb R} \, \ell(D_1 \to y) = \frac{1}{n b} - \log \; \Big((1 - c) + c \exp\left(\frac{1}{n b} \right) \Big).
\end{equation*}

\end{document}